\documentclass[hyper,a4paper]{JHEP3}
\usepackage{epsfig}
\usepackage{latexsym}
\usepackage{graphicx}
\usepackage{amsmath}
\usepackage{amsfonts}
\usepackage{amssymb}
\usepackage{bm}
\usepackage{cite}
\usepackage{dcolumn}
\usepackage{bm}
\usepackage{amssymb}
\usepackage{mathrsfs}
\usepackage{multirow}
\usepackage{amsmath}
\usepackage{slashed}
\newcommand{\mathsym}[1]{{}}

\newcommand{\baz}{\begin{array}{cc}}
\newcommand{\bad}{\begin{array}{ccc}}
\newcommand{\ba}{\begin{array}{c}}
\newcommand{\ea}{\end{array}}
\newcommand{\be}{\begin{equation}}
\newcommand{\ee}{\end{equation}}
\newcommand{\bea}{\begin{eqnarray}}
\newcommand{\eea}{\end{eqnarray}}
\newcommand{\nn}{\nonumber}
\newcommand{\bi}{\begin{itemize}}
\newcommand{\ei}{\end{itemize}}
\newcommand{\bmt}{\begin{pmatrix}}
\newcommand{\emt}{\end{pmatrix}}
\newcommand{\bt}{\begin{tabular}}
\newcommand{\et}{\end{tabular}}

\newcommand{\ovl}{\overline}

\newcommand{\benu}{\begin{enumerate}}
\newcommand{\eenu}{\end{enumerate}}

\newcommand{\bav}{\begin{array}{cccc}}

\def\gs{\mathrel{
   \rlap{\raise 0.511ex \hbox{$>$}}{\lower 0.511ex \hbox{$\sim$}}}}
\def\ls{\mathrel{
   \rlap{\raise 0.511ex \hbox{$<$}}{\lower 0.511ex \hbox{$\sim$}}}}

\numberwithin{equation}{section}

\title{Neutrino masses, dominant neutrinoless double beta decay, and observable lepton flavor violation in left-right
models and SO(10) grand unification with low mass $\bf W_R, Z_R$ bosons}
\author{{\small Ram Lal Awasthi,$^\delta$ ~M. K. Parida$^\dagger$ and ~Sudhanwa Patra$^\dagger$}\\ 
$^\dagger$Center of Excellence in Theoretical and Mathematical Sciences, 
Siksha \textquoteleft O\textquoteright Anusandhan University, Bhubaneswar-751030, India\\
$^\delta$Harish-Chandra Research Institute, Chhatnag Road, Jhusi, Allahabad 211019, India, \\ 
E-mails: \email{ramlal@hri.res.in,  paridamk@soauniversity.ac.in, \\ \quad \quad \quad \quad sudha.astro@gmail.com}}

\abstract{
While the detection of $W_R$-boson at the Large Hadron Collider is likely to resolve 
the mystery of parity violation in weak interaction, observation of neutrinoless double 
beta decay ($0\nu\beta\beta$) is expected to determine whether neutrinos are Majorana fermions. 
 In this work we consider a class of LR models with 
TeV scale $W_R, Z_R$ bosons but having parity restoration at high scales where they originate from well 
known Pati-Salam symmetry or $SO(10)$ grand unified theory minimally extended to accommodate inverse 
seesaw frame work for neutrino masses. Most dominant new contribution to neutrinoless double beta decay 
is noted to occur via $W_L^{-}W_L^{-}$ mediation involving lighter sterile neutrino exchanges. 
The next dominant contribution is found to be through $W_L^{-}W_R^{-}$ mediation involving both light and heavy right-handed 
neutrino or sterile neutrino exchanges. The quark-lepton symmetric origin of the computed value of the Dirac neutrino mass matrix 
is also found to play a crucial role in determining these and other results 
on lepton 
flavor violating branching ratios for $\tau \rightarrow e + \gamma$, $\tau \rightarrow \mu 
+ \gamma$, and $\mu \rightarrow e + \gamma$ accessible to ongoing search experiments. 
The underlying non-unitarity matrix is found to manifest in  substantial 
CP-violating effects even when the leptonic Dirac phase $\delta_{\rm CP} \simeq 0, \pi, 2 \pi$. 
Finally we explore a possible origin of the model in non-supersymmetric SO(10) grand 
unified theory where, in addition to low mass $W_R^\pm$ and $Z_R$ bosons accessible 
to Large Hadron Collider, the model is found to predict observable neutron-antineutron oscillation and 
lepto-quark gauge boson mediated rare kaon decay with $\mbox{Br} \left(K_{\rm L} \rightarrow 
\mu\, \bar{e}\right) \simeq \left(10^{-9}- 10^{-11} \right)$.}
\begin{document}
\section{Introduction}
The Standard Model~(SM) of strong, weak, and electromagnetic interactions has successfully 
confronted numerous experimental tests, yet its failures are exposed in neutrino masses and 
mixings, dark matter, dark energy, and baryon asymmetry of the universe. Apart from having
a number of unknown parameters, the model does not explain why parity violation is monopoly 
of weak interaction. The suggestion of the origin of parity restoration is almost as old as 
the suggestion of parity violation itself when Lee and Yang \cite{yang} conjectured all basic
interactions to be left-right symmetric. Subsequently, two classes of theories have been proposed 
to achieve the desired goal: ({\bf a}) mirror symmetric extension of the Standard Model \cite{mirror}, 
({\bf b}) proposal of left-right symmetric gauge theory based upon {\bf $SU(2)_L \times SU(2)_R \times 
SU(4)_C \left(\equiv \mathcal{G}_{224D} \right) $} \cite{pati} and {\bf $SU(2)_L \times 
SU(2)_R \times U(1)_{B-L} \times SU(3)_C\, \left(\equiv \mathcal{G}_{2213D} \right)$} \cite{moh-senj} 
with $g_{2L} = g_{2R}$. The minimal rank 5 grand unified theory (GUT) like SO(10) \cite{so10} contains 
$G_{224D}$ and $G_{2213D}$ as its subgroups. The canonical ($\equiv$ type-I) and type-II seesaw 
mechanisms  \cite{typeI, valle-typeI, typeII, moh-typeII} explaining tiny left-handed (LH) neutrino masses emerge naturally 
from SO(10), $G_{224D}$, and $G_{2213D}$ gauge theories provided both left-handed (LH) and 
right-handed (RH) neutrinos are Majorana fermions
\be 
m^I_\nu=-M_D\frac{1}{M_N}M_D^T\, ,\quad \quad \quad \,  \quad \quad \quad \quad  m^{II}_\nu=f\,v_L\, ,
\ee   
where $m^I_\nu\, (m^{II}_\nu)$ is the type-I (type-II) prediction of light LH neutrino mass matrix, 
$M_D\, (M_N)$ is the Dirac (right-handed Majorana) neutrino mass. Here 
\be
v_L \simeq \beta\, v^2_{\rm wk}/M_{\Delta_L}\, ,
\label{eq:seesaw}
\ee
$\beta$ is a Higgs quartic coupling, $v_{\rm wk}$ is the electroweak VEV, and $M_{\Delta_L}$ is the left-handed 
triplet Higgs mass. 
Currently a number of dedicated experiments on neutrinoless double beta ($0\nu \beta \beta$) decay 
are in progress \cite{expt-0nu2beta} while Heidelberg-Moscow experiment \cite{claim} has already 
claimed to have measured the effective mass parameter $M^{\rm eff}_{ee} \simeq (0.23-0.56)\,$ eV 
and this observation might be hinting towards the Majorana nature of the light neutrinos \cite{Maj}.

SO(10) GUT has the advantage of unifying the three basic forces (excluding gravity) and all 
fermions of the SM plus the RH neutrinos are unified into its single spinorial representation 
${\bf 16}$. In view of the underlying quark-lepton symmetry $\mathcal{G}_{224D}$ \cite{pati} 
of $SO(10)$, the Dirac neutrino mass matrix $M_D$ could be similar to the up-quark mass matrix 
$M_u$ in these theories, although $\mathcal{G}_{2213D}$ \cite{moh-senj} has also the alternative possibility of 
$M_D$ to be similar to the charged lepton mass matrix $M_\ell$ if the symmetry does not emerge 
from $\mathcal{G}_{224D}$ or $SO(10)$. In any case, if $M_D$ is similar to the up-quark mass matrix or the charged lepton mass matrix, tiny neutrino masses uncovered by the neutrino 
oscillation experiments \cite{DAYA-BAY} push the seesaw scales to be $>10^{10}$~GeV rendering both the seesaw 
mechanisms to be inaccessible for direct experimental tests. In the process, the large scale of the 
associated RH gauge bosons ($W^\pm_R,\, Z_R$) prevent any visible nonstandard impact on weak 
interaction phenomenology including $0\nu\beta\beta$ decay \cite{moh-typeII} while throwing out the origin of 
left-right symmetric theory out of the arena of direct experimental tests at Large Hadron Collider (LHC) and other high energy 
accelerators in foreseeable future. 


In attempts to conventional TeV scale parity-conserving left-right symmetric (LRS) model, it has been shown how 
type-II seesaw formula could be applied for light neutrino masses and mixings  
\cite{tello-senj,Miha} 
and how dominant contribution to neutrino-less double beta  
decay  emerges in the $W_R^{-}-W_R^{-}$ mediated channel. It is expected that this theory could be 
verified by the Large Hadron Collider and other low energy experiments within the next few years.
 
In this work we show how in a different class of LR models\cite{Dparity11} originating from high scale parity restoring Pati-Salam symmetry or SO(10) grand unification, TeV scale $W_R, Z_R$ bosons accessible to LHC are predicted. 
The  Pati-Salam symmetry or $SO(10)$ grand unified theory each are minimally extended with one singlet fermion per generation to accommodate 
the experimentally verifiable gauged inverse seesaw frame work for neutrino masses \cite{inv,inv11,grimus}. Exploiting the other attractive aspect of such quark-lepton unified theories to represent fermion masses,
 we obtain the Dirac neutrino mass matrix as a natural prediction from the GUT-scale fit to all charged fermion masses. 
 The type-I seesaw contribution to neutrino mass cancels out \cite{mass-mixing,majee} and the type-II seesaw contribution and another induced contribution are shown to be subdominant. As a result, the experimentally testable gauged inverse 
 seesaw mechanism \cite{mass-mixing,majee} governs the light neutrino masses.
 The TeV scale masses of $W_R^\pm$ 
and $Z_R$ gauge bosons, and RH neutrinos are  also 
directly accessible to accelerator tests \cite{WR-limit,bound:wr,Kyong,bound:mzr1, bound:mzr2}. 

For the first time we show that this  model originating from high scale quark-lepton symmetry, gives quite dominant new contributions to 
$0\nu\beta\beta$  rate through the $W_L-W_L$ mediation via relatively light sterile neutrino exchange.  The next dominant contribution is found to occur  
 through the  $W_L^{-}-W_R^{-}$ 
mediation with exchanges of light LH and heavy RH neutrinos or sterile neutrinos. The model also gives 
substantial non-unitarity effects and lepton flavor violating (LFV) decays 
accessible to ongoing 
experimental searches for $\tau \rightarrow e + \gamma$, $\tau \rightarrow \mu + \gamma$ 
, and $\mu \rightarrow e + \gamma$. Both the Dirac neutrino mass matrix and the  sterile neutrino masses are found to play significant roles in enhancing the
 $0\nu\beta\beta$ rates in the $W_L-W_L$ channel, LFV
decay branching ratios, and new contributions to CP-violation due to non-unitarity effects.
 
Consistent with current PDG \cite{PDG} values of precision data on electroweak mixing angle ( $\sin^2\theta_W (M_Z)$), the QCD coupling constant ($\alpha_S(M_Z)$), and the electromagnetic fine-structure constant ($\alpha(M_Z)$), while the $SO(10)$ embedding of the conventional TeV scale parity-conserving LRS model has not been possible so far,  we show how the present LR asymmetric  gauge theory 
near the TeV scale with $g_{2L}\neq g_{2R}$ emerges from a non-supersymmetric (non-SUSY) grand unification framework like SO(10) or high scale Pati-Salam symmetry. These two grand 
unified theories also predict observable neutron-antineutron oscillation \cite{nnbar} and
rare kaon decay with $\mbox{Br} \left(K_L\rightarrow \mu\, \overline{e}\right) \simeq \left(10^{-9}- 10^{-11} \right)$ \cite{rarek}
mediated by lepto-quark gauge boson of $SU(4)_{C}$, although proton lifetime is found to be beyond 
the accessible limit of ongoing experiments. The derivation of the Dirac neutrino mass matrix used in all relevant computations is explicitly discussed in the context of high scale Pati-Salam symmetry or SO(10) grand unification.  
 
The plan of this paper is organized as follows: in Sec.\,{\bf 2} we briefly discuss the 
TeV scale left-right gauge theory with low-mass $W_R$, $Z_R$ bosons, light neutrino masses and associated 
non-unitarity effects; in Sec.\,{\bf 3}, we present various Feynman amplitudes for neutrinoless 
double beta decay; in Sec.\,{\bf 4}, we give a detailed discussion for standard and non-standard contributions
to the effective mass parameter for $0\nu 2 \beta$ decay rate and in Sec.\,{\bf 5}, we have discussed the branching 
ratios for lepton flavor violating decays. In Sec.\,{\bf 6}, we implement the idea in 
a SO(10) grand unified theory and derive Dirac neutrino mass matrix at the TeV scale. 
In Sec.\,{\bf 7} we summarize and conclude our results.


\section{Low scale left-right gauge theory and extended seesaw mechanism}
\subsection{The model}
As in the case of extended seesaw mechanism in LR models \cite{majee, mkp-spatra}, besides the standard 16-fermions per generation including the RH neutrino,
, we add one additional sterile fermion singlet for each generation ($S_i, i=1,2,3$).
We start with parity conserving left-right symmetric gauge theory, $\mathcal{G}_{224D}$ \cite{pati} or 
$\mathcal{G}_{2213D}$ \cite{moh-senj}, with equal gauge couplings $(g_{2L}=g_{2R})$ at high scales. In the 
Higgs sector we need both LH and RH triplets $({\bf \Delta_L, \Delta_R})$ as well as the LH 
and RH doublets $\left({\bf \chi}_L, {\bf \large \chi}_R \right)$ in addition to the bidoublet $({\bf \Phi})$ and a 
D-parity odd singlet ${\bf \sigma}$ \cite{Dparity11}. Their transformation properties under $\mathcal{G}_{224D}
\supset \mathcal{G}_{2213D}$ are
\bea
& & \Delta_L(3, 1, 10)\supset \Delta_L(3, 1, -2, 1),\, \,\Delta_R(1, 3, \ovl{10})\supset \Delta_R(1, 3, -2, 1),\nn\\ 
& & \chi_L(2, 1, 4)\supset \chi_L(2, 1, -1, 1), \, \, \chi_R(2, 1, \ovl{4})\supset \chi_R(1, 2, -1, 1) \nn\\
& & \sigma (1, 1, 1)\supset \sigma (1,1,0,1) \nn\\
& & \Phi(2, 2, 1)\supset \Phi(2, 2, 0, 1)\,.
\eea
When the D-parity odd singlet {\bf $\sigma$} acquires a VEV $\langle \sigma\rangle\sim M_P$, the LR discrete symmetry
is spontaneously broken but the gauge symmetry $G_{2213}$ remains unbroken leading to $M^2_{\Delta_R}=(M^2_\Delta-\lambda_\Delta\langle\sigma\rangle M')$, $M^2_{\chi_R}=(M^2_\chi-\lambda_\chi\langle\sigma\rangle M')$,
where $\lambda_\Delta, \lambda_\chi$ are trilinear couplings and $\langle\sigma\rangle$, $M'$, $M_\Delta$, 
$M_\chi$ are all $\sim {\cal O}(M_P)$, the RH Higgs scalar masses are made lighter depending upon the 
degree of fine tuning in $\lambda_\Delta$ and $\lambda_\chi$. The asymmetry in the Higgs sector causes
asymmetry in the $SU(2)_L$ and $SU(2)_R$ gauge couplings with $g_{2L}(\mu)> g_{2R}(\mu)$ for $\mu<M_P$. If one wishes to have $W_R, Z_R$ mass predictions at nearly the same scales and generate Majorana neutrino masses, it is customary to    
 break $G_{2213}\rightarrow{\rm SM}$ by the VEV of the right handed triplet 
$\langle\Delta^0_R\rangle\sim v_R$. We rather suggest a more appealing 
phenomenological scenario with $M_{W_R} > M_{Z_R}$ for which two step breaking of the asymmetric gauge theory to the SM is preferable :
$G_{2213} \stackrel{M^+_R}{\longrightarrow}  G_{2113} \stackrel{M^0_R}{\longrightarrow}  {\rm SM}$, where the first step of breaking 
that generates massive $W_R^{\pm}$ bosons is implemented through the VEV of the heavier triplet $\Sigma_R(1,3,0,1)$ carrying $B-L=0$ 
and the second step of breaking is carried out by $\langle\Delta^0_R\rangle\sim v_R$. At this stage the RH neutral gauge boson gets 
mass which is kept closer to the current experimental lower bound $M_{Z^{\prime}}\ge 1.162$ TeV for its visibility by high energy accelerators. 
We further gauge the extended seesaw mechanism at the TeV scale
 for which the VEV of the RH-doublet $\langle \chi^0_R \rangle=v_\chi$ provides the $N-S$ mixing. The 
$G_{2113}$ symmetric low-scale Yukawa Lagrangian is
\begin{eqnarray}
\mathcal{L}_{\rm Yuk} &= & Y^{\ell} \overline{\psi}_L\, \psi_R\, \Phi 
                       + f\, \psi^c_R\, \psi_R \Delta_R 
                       + F\, \overline{\psi}_R\, S\, \chi_R \nonumber \\
                       &+& S^T \mu_S S +\text{h.c.} 
\end{eqnarray}
which gives rise to the $9\times 9$ neutral fermion mass matrix after electroweak symmetry breaking

\begin{equation}
\mathcal{M}= \left( \begin{array}{ccc}
                0        & 0 & M_D   \\
              0 &    \mu_S         & M \\
              M^T_D & M^T & M_N
                      \end{array} \right) \, ,
\label{eqn:numatrix}       
\end{equation}
where $M_D=Y\langle \Phi\rangle$, $M_N=fv_R$, $M=F\langle\chi_R^0\rangle$. 
It is well known that the mass matrix $M_D$ is determined from high scale symmetry and 
fits to charge fermion masses. In principle the ${\small N-S}$ mixing mass matrix $M$ can assume 
any $3\times 3$, but for the sake of simplicity and economy of parameters we have found 
that the relevant model predictions are possible even if we choose it to have diagonal 
structure. In this case the three diagonal elements can be constrained by the existing 
experimental bound on a unitarity violating parameter. We have also utilized a predicted 
diagonal structure for $M_N$ as well as other diagonal and general forms consistent with 
the $SO(10)$ GUT model. 
\subsection{Neutrino masses and mixings}
In this model the RH neutrinos being heavier than the other mass scales with 
$M_N > M \gg M_D, \mu_S$ are at first integrated out from the Lagrangian leading to \cite{mass-mixing}
\begin{eqnarray}
- \mathcal{L}_{\rm eff} &=& \left(M_D \frac{1}{M_N} M^T_D\right)_{\alpha \beta}\, \nu^T_\alpha \nu_\beta +
\left(M_D \frac{1}{M_N} M^T \right)_{\alpha m}\, \left(\overline{\nu_\alpha} S_m + \overline{S_m} \nu_\alpha \right)
\nonumber \\
&&\hspace*{4.0cm} +\left(M \frac{1}{M_N} M^T\right)_{m n}\, S^T_m S_n - \mu_S S^T_m S_n \, ,
\end{eqnarray}
which, in the $\left(\nu,~ S\right)$ basis, gives the $6 \times 6$ mass matrix
\begin{eqnarray}
\mathcal{M}_{\rm eff} =- \left( \begin{array}{cc}
    M_DM_N^{-1} M^T_D  &  M_D M_N^{-1} M^T    \\
    MM_N^{-1} M_D^T       & MM_N^{-1} M^T - \mu_S 
        \end{array} \right) \, ,
\label{eqn:eff_numatrix}       
\end{eqnarray}
while the $3\times 3$ heavy RH neutrino mass matrix $M_N$ is the other part of the full 
$9 \times 9$ neutrino mass matrix. This $9 \times 9$ mass matrix $\tilde{\mathcal{M}}_{\rm 
\tiny BD}$ which results from the first step of block diagonalization procedure as discussed 
above and in the Appendix A is
\begin{eqnarray}
\mathcal{W}^\dagger_1 \mathcal{M}_\nu \mathcal{W}^*_1 = 
\tilde{\mathcal{M}}_{\rm \tiny BD}
 = \bmt \mathcal{M}_{\rm eff} & 0\\
0& M_N
\emt \, ,
\end{eqnarray}
where $\mathcal{W}_1$ has been derived as shown in eqn.\,(\ref{app:w1}) of 
Appendix A.

\noindent
After the second step of block 
diagonalization, the type-I seesaw contribution cancels out and gives in the $\left(\nu, S, N \right)$ basis
\begin{eqnarray}
\mathcal{W}^\dagger_2 \tilde{\mathcal{M}}_{\rm \tiny BD} \mathcal{W}^*_2 = \mathcal{M}_{\rm \tiny BD}
= \bmt m_\nu &0&0\\
0&m_{\cal S}&0\\
0&0&m_{\cal N}
\emt\, ,
\label{eqn:block-form}    
\end{eqnarray}
where $\mathcal{W}_2$ has been derived in eqn.\,(\ref{app:w2}) of the appendix. 
In eqn. (\ref{eqn:block-form}), the three $3 \times 3$ matrices are 
\bea
m_{\nu} & \sim &  M_D M^{-1} \mu_S  (M_DM^{-1})^T  \\ 
m_{\cal S}& \sim & \mu_S - M M^{-1}_NM^T  \\ 
m_{\cal N} & \sim &  M_N \, ,
\label{eq:mass}
\eea
the first of these being the well known inverse seesaw formula \cite{inv,inv11}.

In the third step, $m_{\nu}$, $m_{\cal S}$, and $m_{\cal N}$ are further diagonalized 
by the respective unitary matrices to give their corresponding eigenvalues
\begin{eqnarray}
U^\dagger_\nu\, m_{\nu}\, U^*_{\nu}  &=& \hat{m}_\nu = 
         \text{diag}\left(m_{\nu_1}, m_{\nu_2}, m_{\nu_3}\right)\, , \nonumber \\ 
U^\dagger_S\, m_{\cal S}\, U^*_{S}  &=& \hat{m}_S = 
         \text{diag}\left(m_{S_1}, m_{S_2}, m_{S_3}\right)\, , \nonumber \\
U^\dagger_N\, m_{\cal N}\, U^*_{N}  &=& \hat{m}_N = 
         \text{diag}\left(m_{N_1}, m_{N_2}, m_{N_3}\right)\, .
\label{eq:nudmass}
\end{eqnarray}
\noindent
The complete mixing matrix \cite{grimus, mkp-spatra} diagonalizing the above $9 \times 9$ 
neutrino mass matrix given in eqn. 
(\ref{eqn:numatrix}) turns out to be
 \begin{eqnarray}
\mathcal{V}&\equiv&
\bmt 
{\cal V}^{\nu\hat{\nu}}_{\alpha i} & {\cal V}^{\nu{\hat{S}}}_{\alpha j} & {\cal V}^{\nu \hat{N}}_{\alpha k} \\
{\cal V}^{S\hat{\nu}}_{\beta i} & {\cal V}^{S\hat{S}}_{\beta j} & {\cal V}^{S\hat{N}}_{\beta k} \\
{\cal V}^{N\hat{\nu}}_{\gamma i} & {\cal V}^{N\hat{S}}_{\gamma j} & {\cal V}^{N\hat{N}}_{\gamma k} 
\emt \\
&=&\bmt 
\left(1-\frac{1}{2}XX^\dagger \right) U_\nu  & 
\left(X-\frac{1}{2}ZY^\dagger \right) U_{S} & 
Z\,U_{N}     \\
-X^\dagger\, U_\nu   &
\left(1-\frac{1}{2} \{X^\dagger X + YY^\dagger \}\right) U_{S} &
\left(Y-\frac{1}{2} X^\dagger Z\right) U_{N}   \\
y^*\, X^\dagger\, U_{\nu} &
-Y^\dagger\, U_{S} &
\left(1-\frac{1}{2}Y^\dagger Y\right)\, U_{N} 
\emt \, ,
 \label{eqn:Vmix-extended}
\end{eqnarray}
as shown in the appendix. In eqn.\,(\ref{eqn:Vmix-extended}) $X = M_D\,M^{-1}$, $Y=M\, 
M^{-1}_N$, $Z=M_D\,M^{-1}_N$, and $y=M^{-1}\,\mu_S$.
\subsection{The unitarity violating matrix}
In this subsection we discuss briefly how non-unitarity arises in the lepton sector and how existing bounds 
on lepton flavor violating processes impose upper bounds on all the elements of the $3\times 3$ non-unitarity 
matrix $\eta$.  We have shown in Sec.\,6 how different forms of the $3\times3$ RH neutrino mass matrix $M_N$ 
are allowed by the fermion mass fits at the GUT scale including a diagonal form with specific eigen values. 
These matrices have been used to estimate model predictions in Sec.\,4-Sec.\,6. Using the constrained diagonal 
form of $M$ as mentioned above, the mass matrix $\mu_S$ is determined using the gauged inverse see-saw formula 
and neutrino oscillation data provided that the Dirac neutrino mass matrix $M_D$ is also known. The determination 
of $M_D$ at the TeV scale, basically originating from high-scale quark-lepton symmetry $G_{224D}$ or $SO(10)$ GUT, 
is carried out by predicting its value at the high scale from fits to the charged fermion masses of three generations 
and then running down to the lower scales using the corresponding renormalization group equations (RGEs) in the 
top-down approach. It is to be noted that for fits to the fermion masses at the GUT scale, their experimental 
values at low energies are transported to the GUT scale using RGEs and the bottom-up approach. This procedure 
has been carried out in Sec.\,6.5 by successfully embedding the LR gauge theory in a suitable non-SUSY $G_{224}$ 
and $SO(10)$ framework and the result is
 
\begin{eqnarray}
\label{eq:md_with_rge}
M_D =  \left( \begin{array}{ccc} 
 0.02274          & 0.09891-0.01603i    & 0.1462-0.3859i\\
 0.09891+0.01603i & 0.6319              & 4.884+0.0003034i\\
 0.1462+0.3859i   & 4.884-0.0003034i    & 117.8
\end{array} \right) \text{GeV}.\,
\end{eqnarray}
This value of $M_D$  will be utilized for all applications discussed subsequently in this work including 
the fit to the neutrino oscillation data through the inverse seesaw formula, predictions of effective mass 
parameters in $0\nu\beta\beta$, computation of non-unitarity and CP-violating effects, and LFV decay branching ratios. 

Usually diagonalizing the light active Majorana neutrino mass matrix by the PMNS mixing matrix $U_{\rm PMNS}$ gives
$U_{\rm PMNS}^{\dagger}\, m_{\nu}\, U^*_{\rm PMNS} = \hat{m}_{\nu} = \text{diag}\left(m_{\nu_1},m_{\nu_2},m_{\nu_3}\right)$.
But, in this extended seesaw scheme, diagonalization is done by a matrix ${\cal N\equiv V^{\nu\nu}}$ which is a 
part of the full $9\times 9$ mixing matrix ${\cal V}_{9\times 9}$. Using $\eta=\frac{1}{2}XX^\dagger$ where $X=M_D/M$, this 
diagonalizing matrix is
\be 
{\cal N}\simeq(1-\eta)U_{\rm PMNS}.
\ee 
\begin{table}[b]
\centering
\begin{tabular}{|c|c|c|c|c|}
\hline\hline 
{\bf \small measure of}    & Expt. bound \cite{antbnd} &  &  & \\ 
{\bf non-unitarity}        & {\bf C0} & {\bf C1} & {\bf C2} & {\bf C3}\\ \hline
${\bf |\eta_{ee}| }$       &$2.0\times 10^{-3}$&$3.5\times 10^{-8}$&$2.7\times 10^{-7}$ &$3.1\times 10^{-6}$  \\ 
${\bf |\eta_{e\mu}|}$      &$3.5\times 10^{-5}$&$3.9\times 10^{-7}$&$3.4\times 10^{-6}$&$1.5\times 10^{-5}$\\ 
${\bf |\eta_{e\tau}|}$     &$8.0\times 10^{-3}$&$9.4\times 10^{-6}$&$2.8\times 10^{-5}$&$6.4\times 10^{-5}$\\ 
${\bf |\eta_{\mu\mu}|}$    &$8.0\times 10^{-4}$&$4.7\times 10^{-6}$&$2.3\times 10^{-5}$&$6.9\times 10^{-5}$\\
${\bf |\eta_{\mu\tau}|}$   &$5.1\times 10^{-3}$&$1.1\times 10^{-4}$&$2.2\times 10^{-4}$&$3.2\times 10^{-4}$\\
${\bf |\eta_{\tau\tau}|}$  &$2.7\times 10^{-3}$&$2.7\times 10^{-3}$&$2.7\times 10^{-3}$&$2.7\times 10^{-3}$\\
\hline \hline 
\end{tabular}
\caption{Experimental bounds of the non-unitarity matrix elements $|\eta_{\alpha\beta}|$ (column ${\bf C0}$) 
         and their predicted values for degenerate (column ${\bf C1}$), partially-degenerate (column ${\bf C2}$), 
         and non-degenerate (column ${\bf C3}$) values of  $M=\mbox{diag} \left(M_1, M_2, M_3\right)$ as described 
         in cases (${\bf a}$), (${\bf b}$) and (${\bf c}$), respectively, in the text.}
\label{tab:eta-bound}
\end{table}

Thus $\eta$ is a measure of deviation from unitarity in the lepton sector on which there has been 
extensive investigations in different models \cite{ramlal, mkp-spatra, antbnd, Bdev-non, non-unit}. Assuming $M$ to be diagonal 
for the sake of simplicity, $M\equiv{\rm diag}(M_1, M_2, M_3)$, gives $\eta_{\alpha\beta}=
\frac{1}{2} \sum_{k}\,{M_D}_{\alpha k}M^{-2}_k{M^*_D}_{\beta k}$, but it can be written explicitly for the degenerate case 
($M_1=M_2=M_3=M_d$)
\begin{eqnarray} 
\label{eq:input-eta-ckm}
\eta = \frac{\text{1 GeV}^2}{M^2_d (\text{GeV}^2)} 
 \left( \begin{array}{ccc} 
0.0904          & 0.3894-0.9476i   & 8.8544-22.7730i\\ 
0.3894+0.9476i  & 12.1314          & 289.22+0.00005i\\ 
8.8544+22.7730i & 289.22-0.00005i  & 6950.43
\end{array} \right)\, .
\end{eqnarray}

\begin{figure}[h!]
\centering
\includegraphics[scale=.45,angle=-90]{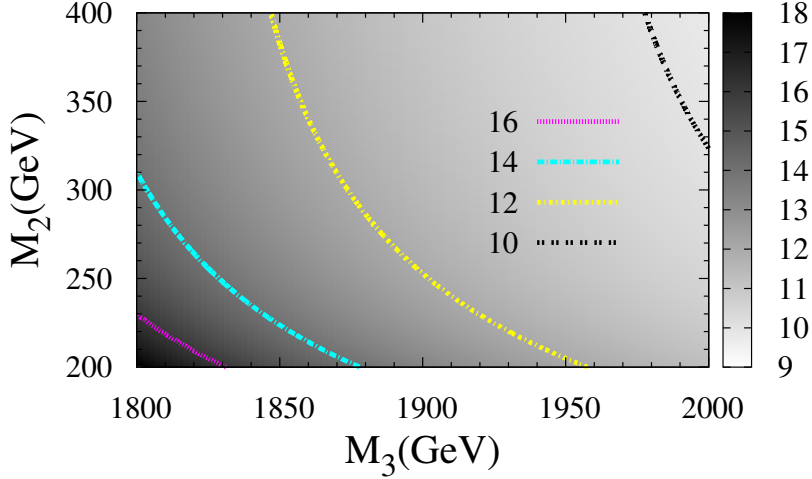}
\caption{The contours of $M_1$ in the plane of $M_2$ and $M_3$. The solid curves 
        in the diagram represent $M_3$ dependence of $M_2$ for fixed values of $M_1$ 
        using eqn.~({\protect\ref{eq:eta_unitarity_bound}}). The brightest top-right 
        corner suggests that lightest $M_1$ may exist for largest values of $M_2$ and $M_3$.}
\label{fig:contour}
\end{figure}
\noindent For the non-degenerate diagonal matrix $M$, saturating the experimental 
bound for $|\eta_{\tau\tau}|<2.7\times 10^{-3}$ \cite{ramlal,mkp-spatra} gives
\begin{eqnarray}
\frac{1}{2}\, \bigg[\frac{0.170293}{M^2_1}+\frac{23.8535}{M^2_2} 
                    +\frac{13876}{M^2_3} \bigg] = 2.7 \times 10^{-3}\, ,
\label{eq:eta_unitarity_bound}
\end{eqnarray}
\noindent
where the three numbers inside the square bracket are in $\mbox{GeV}^2$. The correlation between $M_2$ 
and $M_3$ is shown in Fig.\,{\bf\ref{fig:contour}} where the allowed region in the brightest top right 
corner suggests the possibility of lightest $M_1$ for large values of $M_2$ and $M_3$. It is clear from 
eq.(\ref{eq:eta_unitarity_bound}) that $M_i$ can not be arbitrary. Rather they are ordered with $M_3 > 
M_2 > M_1$ and also they are bounded from below with $M_1 > 5.6$~ GeV, $M_2> 66.4$~ GeV, $M_3 > 1.6 $~ TeV. 
In the degenerate case $M_1=M_2=M_3=1604.4$ GeV. If we assume equal contribution to nonunitarity from 
all three terms in the left hand side of eq.({\ref{eq:eta_unitarity_bound}}), we get $M={\rm diag}(9.7, 
115.1, 2776.6)$~GeV. Besides these constraints, we have used the primary criteria $M_N > M >> M_D,\mu_S$ 
where $M_N \le O(v_R)$, the $G_{2113}$ breaking scale in choosing the elements of $M$.      

The elements of $\eta$ have been listed in the Table.\,\ref{tab:eta-bound} for $\bf (a)$ degenerate 
$M={\rm diag}$(1604.4, 1604.4, 1604.4)~GeV, ${\bf (b)}$ partially degenerate $M={\rm diag}(100, 100, 
2151.58)$~GeV, and ${\bf (c)}$ non degenerate $M={\rm diag}(9.73, 115.12, 2776.6)$~GeV in columns 
${\bf C1, C2}$ and ${\bf C3}$, respectively, where in column ${\bf C0}$, experimental bounds are 
presented \cite{antbnd}. 

\begin{table}[t]
\centering
\begin{tabular}{|l|l|c|}
\hline\hline 
$m_\nu$& $M$ & $\mu_S$\\
\hline
\multirow{3}{*}{} &(a)  & 
$\left(\begin{array}{ccc}
 9.457+4.114i   & -2.073-0.904i    & 0.087-0.001i \\
 . & 0.455+0.198i & -0.019-0.0003i \\
 . & .            & 0.00069-0.000027i
\end{array}\right)${\rm GeV}\\ \cline{2-3}
{NH} &(b)  &$\left(
\begin{array}{ccc}
 0.037+0.016i      & -0.008-0.003i & 0.007-0.0001i \\
 .                 & 0.001+0.0007i & -0.001+0.00002i \\
 .                 & .             & 0.001-0.0004i
\end{array}
\right)${\rm GeV}  \\ \cline{2-3}
 & (c)  & $\left(
\begin{array}{ccc}
 3.476+1.512i  & -9.018-3.933i  & 9.180+0.141i \\
 .             & 23.410+10.230i & -23.840-0.385i \\
 .             & .              & 20.670-8.246i
\end{array}
\right)\times 10^{-4}${\rm GeV} \\ \hline
\hline 
\end{tabular}
\caption{Structure of $\mu_S$ from neutrino oscillation data for normal-hierarchy~(NH) of light neutrino masses, 
$m_\nu = (0.00127, 0.008838,0.04978)$~eV and different 
mass pattern of $M$: ${\bf (a)}$ $M=(1604.442, 1604.442, 1604.442)$~GeV, 
${\bf (b)}$ $M=(100.0, 100.0, 2151.5)$~GeV, and ${\bf (c)}$ $M=(9.72, 115.12, 2776.57)$~GeV.}
\label{tab:mus-NH}
\end{table}
\subsection{Determination of $\mu_S$ from fits to neutrino oscillation data}
We utilize the central values of parameters obtained from recent global 
fit to the neutrino oscillation data \cite{fogli}
\begin{eqnarray}
&&\sin^2\theta_{12}=0.320,\,\,\,\, \sin^2\theta_{23}=0.427,\nonumber \\
&&\sin^2\theta_{13}=0.0246,\,\, \delta_{CP}=0.8 \pi,\nonumber \\
&&\Delta m_{\rm sol}^2=7.58\times 10^{-5}{\rm eV}^2, \nonumber\\
&&|\Delta m_{\rm atm}|^2=2.35\times 10^{-3}{\rm eV}^2,
\end{eqnarray}
and ignore Majorana phases 
$(\alpha_1=\alpha_2=0)$. Then using the non-unitarity mixing matrix ${\cal N}=(1-\eta)\, 
U_{\rm PMNS}$ and the relation $m_\nu={\cal N}\hat{m}_\nu{\cal N}^T$, we derive the 
$\mu_S$ matrix by inverting the inverse seesaw formula,

\begin{table}[htb!]
\centering
\begin{tabular}{|l|l|c|}
\hline\hline 
$m_\nu$&$M$&$\mu_S$\\
\hline
\multirow{3}{*}{} & (a)  &$ \left(
\begin{array}{ccc}
 82.04+2.261i    & -17.75-0.508i    & 0.642-0.251i \\
 .               & 3.842+0.114i     & -0.138+0.054i \\
 .               & .                & 0.0042-0.0040i
\end{array}
\right)${\rm GeV} \\ \cline{2-3}
{IH} & (b)  &$\left(
\begin{array}{ccc}
 0.318+0.0088i   & -0.0689-0.0019i  & +0.0536-0.0209i \\
 .               & +0.0149+0.00044i & -0.0116-0.0045i \\
 .               & .                & 0.0075-0.0073i
\end{array}
\right)${\rm GeV}  \\ \cline{2-3}
 & (c)  &$\left(
\begin{array}{ccc}
 3.015+0.083i    & -7.72-0.221i    & 6.73-2.62i \\
 .               & 19.78+0.58i     & -17.25+6.714i \\
 .               & .               & 12.41-12.08i
\end{array}
\right)\times 10^{-3}${\rm GeV}  \\ \hline
\hline 
\end{tabular}
\caption{Same as Tab.~{\protect\ref{tab:mus-NH}} but for inverted-hierarchy~(IH) 
of light neutrino masses $m_\nu=(0.04901, 0.04978,0.00127)$~eV.}
\label{tab:mus-IH}
\end{table}

\begin{eqnarray}
\label{eq:mu_S_matrix}
\mu_S &=& X^{-1}\, \mathcal{N} \hat{m}_\nu \mathcal{N}^T\, (X^T)^{-1} \nonumber \\
&=&{\small 
\left(
\begin{array}{ccc}
 3.476+1.512i  & -9.018-3.933i  & 9.180+0.141i \\
 .             & 23.410+10.230i & -23.840-0.385i \\
 .             & .              & 20.670-8.246i
\end{array}
\right)\times 10^{-4}\, \mbox{GeV} 
}
\end{eqnarray}
where we have used normal hierarchy (NH) for light neutrino masses, $\hat{m}_{\nu} =(0.00127, 0.00885,$
$0.0495)$~eV in the non degenerate case of $M=\mbox{diag}(9.72, 115.12, 2776.57)$\, GeV. 
For the sake of completeness, we have presented few solutions of $\mu_S$ matrix for degenerate, 
partially-degenerate and non-degenerate values of $M$ as shown in the Tables {\ref{tab:mus-NH} and 
\ref{tab:mus-IH} corresponding to NH and IH light neutrino masses, respectively. For the quasi-degenerate 
(QD) pattern of light neutrino masses the matrix $\mu_S$ can be easily derived and all our 
analyses carried out in Sec.\,{\bf 3} - Sec.{\bf 5} can be repeated. 

\section{Amplitudes for $0\nu\beta\beta$ decay and effective mass parameters}
In this section we discuss analytically the contributions of various Feynman diagrams in $W_L^--W_L^-$ channel 
(with two left-handed currents), $W_R^--W_R^-$ channel (with two right-handed currents), and $W_L^--W_R^-$ channel 
(with one left-handed and one right-handed current) and estimate the corresponding amplitudes in the TeV scale 
asymmetric left-right gauge theory with extended seesaw mechanism.

The charged current interaction Lagrangian for leptons in this model in the flavor basis is 
\begin{eqnarray}
\mathcal{L}_{\rm CC} &=& \frac{g}{\sqrt{2}}\, \sum_{\alpha=e, \mu, \tau}
\bigg[ \overline{\ell}_{\alpha \,L}\, \gamma_\mu {\nu}_{\alpha \,L}\, W^{\mu}_L 
      + \overline{\ell}_{\alpha \,R}\, \gamma_\mu {N}_{\alpha \,R}\, W^{\mu}_R \bigg] + \text{h.c.} 
\label{eqn:ccint-flavor}
\end{eqnarray}
Following the masses and mixing for neutrinos in the extended seesaw scheme \cite{mkp-spatra} 
discussed in Sec.\,{\bf 2}, LH and RH neutrino flavor states are expressed in terms of mass eigenstates 
($\hat{\nu}_i$, $\hat{S}_i$, $\hat{N}_i$)
\begin{eqnarray}
& &\nu_{\alpha\,L} \sim \mathcal{V}^{\nu\nu}_{\alpha\, i}\, \hat{\nu}_i + \mathcal{V}^{\nu\, S}_{\alpha\, i}\, \hat{S}_i +
                      \mathcal{V}^{\nu\, N}_{\alpha\, i}\, \hat{N}_i,  \\
& &N^C_{\alpha\,R} \sim \mathcal{V}^{N\, \nu}_{\alpha\, i}\, \hat{\nu}_i + \mathcal{V}^{N\, S}_{\alpha\, i}\,\hat{S}_i +
                      \mathcal{V}^{NN}_{\alpha\, i}\, \hat{N}_i.
\label{eqn:nustate-mass}
\end{eqnarray}
In addition, there is a possibility where left-handed and right-handed gauge bosons mix with each 
other and, hence, the physical gauge bosons are linear combinations of $W_L$ and $W_R$ as
\begin{equation}
\label{eqn:LRmix}
\left\{ 
\begin{array}{l} 
W_1 = \phantom{-}\cos\zeta_{\rm {\small LR}} ~W_L + \sin\zeta_{\rm {\small LR}} ~W_R \\ 
W_2 = - \sin\zeta_{\rm {\small LR}} ~W_L +  \cos\zeta_{\rm {\small LR}} ~W_R 
\end{array} \right. 
\end{equation}
with 
\begin{equation}
|\tan\, 2\zeta_{\rm \small LR}| \sim \frac{v_u\, v_d}{v^2_R} \sim \frac{v_d}{v_u} \frac{g^2_{\rm 2R}}{g^2_{\rm 2L}} 
\left (\frac{M^2_{W_L}}{M^2_{W_R}} \right) \leq 10^{-4}.
\end{equation}
As it is evident from the charged-current interaction given in eqn.\,(\ref{eqn:ccint-flavor}) and taking left- and right-handed 
gauge boson mixings into account given in eqn.\,(\ref{eqn:LRmix}), 
there can be several Feynman diagrams which contribute to neutrinoless double beta decay transition 
in the TeV scale left-right gauge theory. They can be broadly classified as due to 
$W^-_L - W^-_L$ mediation purely due to two left-handed currents, $W^-_R - W^-_R$ mediation purely 
due to two right-handed currents, and $W^-_L - W^-_R$ mediations due to one left-handed current and 
one right-handed current which are denoted by ${\rm 
LL}$, ${\rm RR}$, and ${\rm LR}$ in the superscripts of the corresponding amplitudes. These 
diagrams are shown in Fig.\,\ref{fig:inverse-WLWL} - Fig.\,\ref{fig:inverse-WLWR}. 
\begin{figure}[htb!]
\includegraphics[width=15cm,height=5cm]{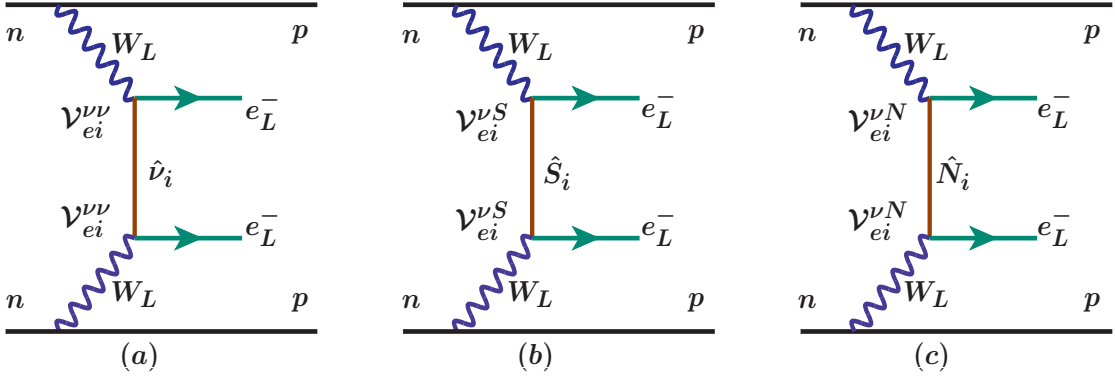}
 \caption{Feynman diagrams for neutrinoless double beta decay ($0\, \nu\, \beta \beta$) contribution 
with virtual Majorana neutrinos $\hat{\nu}_i$, $\hat{S}_i$, and $\hat{N}_i$ along with the mediation of 
two $W_L$-bosons.}
\label{fig:inverse-WLWL}
\end{figure}

\subsection{$W^-_L - W^-_L$ mediation}
The most popular standard contribution is due to $W^-_L - W^-_L$ mediation 
by light neutrino exchanges. But one of our major contribution in this work 
is that even with $W^-_L - W^-_L$ mediation, the sterile neutrino exchange 
allowed within the extended seesaw mechanism of the model can yield much more dominant 
contribution to $0\nu \beta \beta$ decay rate than the standard one. With 
the exchange of left-handed light neutrinos ($\hat{\nu}_i$), sterile neutrinos 
($\hat{S}_j$), and RH heavy Majorana neutrinos ($\hat{N}_k$), the diagrams shown 
in Fig. \ref{fig:inverse-WLWL}.(a), Fig. \ref{fig:inverse-WLWL}.(b), and Fig. 
\ref{fig:inverse-WLWL}.(c) contribute
\begin{eqnarray}
\label{eq:amp_ll} 
& &\mathcal{A}^{LL}_{\nu} \propto \frac{1}{M^4_{W_L}} \sum_{i=1,2,3} \frac{\left(\mathcal{V}^{\nu \nu}_{e\,i}\right)^2\, m_{\nu_i}}{p^2} \,,\\
& &\mathcal{A}^{LL}_{S} \propto \frac{1}{M^4_{W_L}} \sum_{j=1,2,3} \frac{\left(\mathcal{V}^{\nu S}_{e\,j}\right)^2}{m_{S_j}} \, ,  \\ 
& &\mathcal{A}^{LL}_{N} \propto \frac{1}{M^4_{W_L}} \sum_{k=1,2,3} \frac{\left(\mathcal{V}^{\nu N}_{e\,k}\right)^2}{m_{N_k}} \, ,
\end{eqnarray}
where $|p^2| \simeq \left(190~\mbox{MeV}\right)^2$ represents neutrino virtuality momentum 
\cite{nuvirt-moh}.\\

\begin{figure}[htb!]
\centering
\includegraphics[scale=0.85]{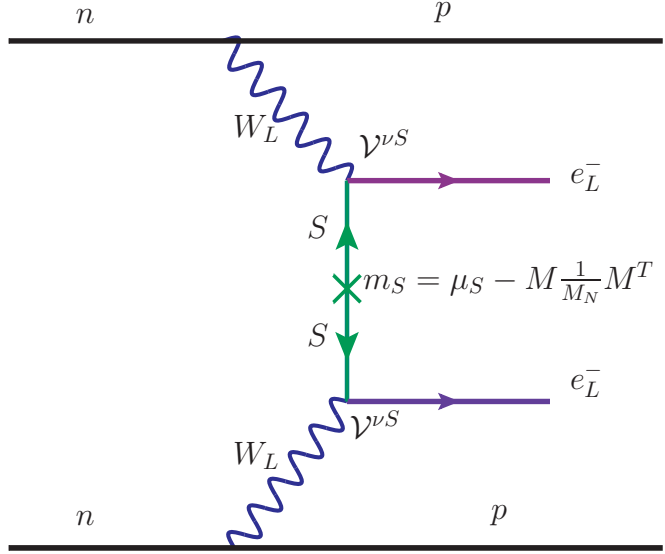}
 \caption{Feynman diagram for neutrinoless double beta decay  contribution by $W_L^--W_L^-$ 
          mediation and by the exchange of virtual sterile neutrinos (${S}$). The Majorana 
          mass insertion has been shown explicitly by a cross.}
\label{fig:sterile-WLWL}
\end{figure}

To understand the origin and the role of the relevant Majorana mass insertion terms as source of 
$|\Delta L|=2$ lepton number violation in the new contribution to $0\nu\beta\beta$ process, we 
briefly discuss the example of sterile fermion ($S$) exchange corresponding to Fig.\,\ref{fig:inverse-WLWL}.(b) 
and Fig.\,\ref{fig:sterile-WLWL}. At first we note that, in contrast to the inverse seesaw framework 
with pseudo-Dirac type RH neutrinos \cite{Bdev-non,ramlal} where the only source of $|\Delta L|= 2$ 
lepton number violation is $\mu_S$, in the present case of extended seesaw the Majorana mass for $S$ 
gets an additional dominant contribution $M M_N^{-1}M^T$ as shown explicitly in eq.\,(2.4) and eq.\,(2.9). 
The expanded form of the Feynman diagram with both the mass insertion terms is shown in Fig.\,\ref{fig:sterile-WLWL} 
which gives
\be 
\mathcal{A}^{LL}_{S} \propto \frac{1}{M^4_{W_L}}\, P_L\bigg[\mathcal{V}^{\nu S}
\frac{1}{p\hspace{-0.2 cm}\slash -m_S}m_S\frac{1}{p\hspace{-0.2 cm}\slash -m_S}{\mathcal{V}^{\nu S^T}}\bigg]_{\rm ee} P_L, \label{addedeqn1}
\ee
where we have used $m_{S} =  \mu_S - M\,M^{-1}_N\,M^T$. Within the model approximation and allowed 
values of parameters, $|m_S| \simeq |M\,M_N^{-1}\,M^T| \gg |p| \gg |\mu_S| $ resulting in
\be
\mathcal{A}^{LL}_{S} \propto \frac{1}{M^4_{W_L}}{\bigg[\mathcal{V}^{\nu S} \left(
\frac{\mu_S}{m_S^2}+\frac{1}{m_S}\right){\mathcal{V}^{\nu S^T}}\bigg]
}_{\rm ee}, \label{addedeqn2}
\ee
where the first term is negligible compared to the second term, and we get eq.\,(3.7). On the other hand, 
in the case of pseudo-Dirac RH neutrinos corresponding to $M_N=0$ in eq.\,(2.3), the only Majorana mass 
insertion term in Fig.\,\ref{fig:sterile-WLWL} is through $m_S=\mu_S$ with $|\mu_S| \ll |p|$. Then 
eq.\,(\ref{addedeqn1}) gives $\mathcal{A}^{LL}_{S} \propto \frac{1}{M^4_{W_L}}\, 
\frac{\left(\mathcal{V}^{\nu S}\right)^2\,\mu_S}{p^2} \simeq \frac{1}{M^4_{W_L}}\, \frac{m_\nu}{p^2}$ 
which is similar to the standard contribution. This latter situation is never encountered in the parameter 
space of the present models.

\begin{figure}[htb!]
\includegraphics[width=15cm,height=5cm]{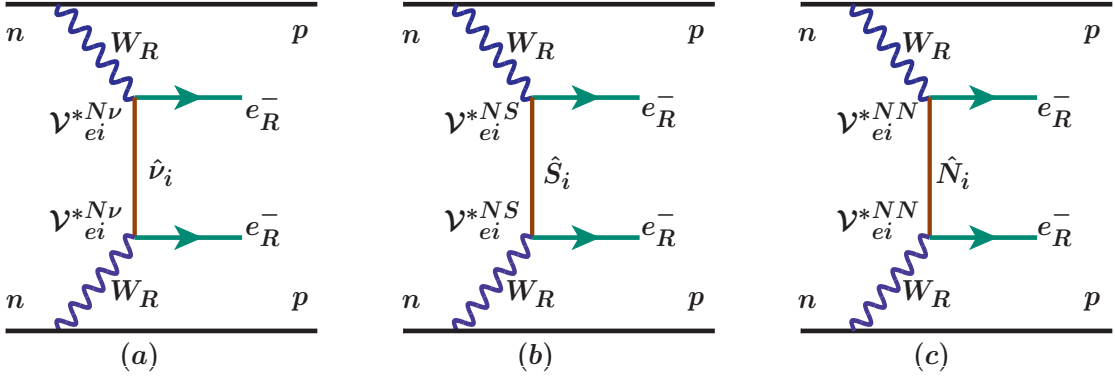}
\caption{Same as Fig.~{\protect\ref{fig:inverse-WLWL}} but with $W_R-W_R$ mediation.}
\label{fig:inverse-WRWR}
\end{figure}
\subsection{$W^-_R - W^-_R$ mediation} 
This contribution arising purely out of right-handed weak currents can also occur 
by the exchanges of $\hat{\nu}_i$, $\hat{S}_i$, and $\hat{N}_i$ and the corresponding diagrams 
are shown in Fig. \ref{fig:inverse-WRWR}.(a), Fig. \ref{fig:inverse-WRWR}.(b), 
and Fig. \ref{fig:inverse-WRWR}.(c) leading to the amplitudes 
\begin{eqnarray}
& &\mathcal{A}^{RR}_{\nu} \propto \frac{1}{M^4_{W_R}} \frac{\left(\mathcal{V}^{N \nu}_{e\,i}\right)^2\, m_{\nu_i}}{p^2} \,, \\ 
& &\mathcal{A}^{RR}_{S} \propto \frac{1}{M^4_{W_R}} \frac{\left(\mathcal{V}^{N S}_{e\,j}\right)^2}{m_{S_j}} \, , \\ 
& &\mathcal{A}^{RR}_{N} \propto \frac{1}{M^4_{W_R}} \frac{\left(\mathcal{V}^{N N}_{e\,j}\right)^2}{m_{N_k}}\, .
\label{eq:amp_rr}
\end{eqnarray}

\begin{figure}[h!]
\centering
\includegraphics[width=12cm,height=4.8cm]{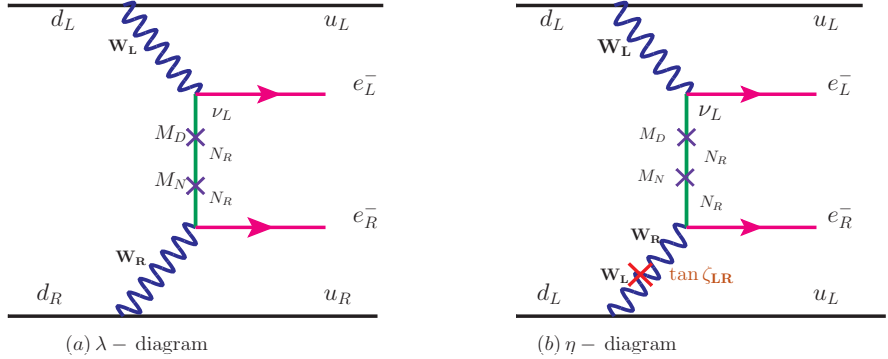}
\caption{Mixed Feynman diagram with $W_L-W_R$ mediation; left-panel is for $\lambda$-mechanism and right-panel 
         is for $\eta$-mechanism as defined in Ref. \cite{doi-kotani} and discussed in the text.}
\label{fig:inverse-WLWR}
\end{figure}
\subsection{$W^-_L - W^-_R$ mediation} 
According to our observation, although these contributions arising out of mixed effects 
by the exchanges of light LH and heavy RH neutrinos and also by the exchange of sterile neutrinos 
are not so dominant compared to those due to $W^-_L - W^-_L$ mediation with sterile neutrino exchanges, 
as discussed in Sec.\,{\bf 3.1}, the amplitudes are stronger than the standard one. The two types of  
mixed helicity Feynman diagrams \cite{vergados, doi-kotani, bery}; 
{\bf (i).} $\underline{\lambda-\mbox{mechansim:}}$ coming from one left-handed and one right-handed current 
($W_L$-$W_R$ mediation) shown in Fig. \ref{fig:inverse-WLWR}.(a), 
{\bf (ii).} $\underline{\eta-\mbox{mechansim:}}$ arising because of additional possibility of $W_L$-$W_R$ 
mixing even though two hadronic currents are left-handed, as shown in Fig. \ref{fig:inverse-WLWR}.(b), leading to a suppression factor 
$\tan \zeta_{LR}$. The corresponding Feynman amplitudes for these mixed helicity diagrams are given below  
\bea
& & \mathcal{A}^{LR}_{\lambda} \propto \frac{1}{M^2_{W_L}\,M^2_{W_R}} 
\left(U_{\nu} \right)_{ei} \left(\frac{M_D}{M_N}\, \right)_{ei}\, \frac{1}{|p|} \, , \\
& & \mathcal{A}^{LR}_{\eta} \propto \frac{\bf \tan \zeta_{LR}}{M^4_{W_L}} 
\left(U_{\nu} \right)_{ei} \left(\frac{M_D}{M_N}\,\right)_{ei}\, \frac{1}{|p|}
\label{eq:amp_lr}
\eea

\subsection{Doubly Charged Higgs contribution} 
Although we have ignored contributions due to exchanges of LH (RH) doubly 
charged Higgs bosons $\Delta^{--}_L$ ($\Delta^{--}_R$) in this work, we 
present the corresponding amplitudes for the sake of completeness,
\begin{enumerate} 
\item 
[(i)] ~~$\mathcal{A}^{LL}_{\Delta_L} \propto \frac{1}{M^4_{W_L}} 
\frac{1}{M_{\Delta_L}^2} f_L v_L $ \, ,

\item
[(ii)]~~
$ \mathcal{A}^{RR}_{\Delta_R} \propto \frac{1}{M^4_{W_R}}
\frac{1}{M^2_{\Delta_{R}}} f_R v_R $\, .
\end{enumerate}
As stated in Sec.\,{\bf 2}, the masses of $\Delta^{--}_L$ and $\Delta^{++}_L$ 
are of the order of the large parity restoration scale which damps out the 
induced VEV ${\it v}_L$ and the corresponding amplitude. The amplitude due 
to $\Delta^{--}_R$ exchange is damped out compared to the standard amplitude 
as it is $\propto \frac{1}{M^5_{W_R}}$.

\subsection{Nuclear matrix elements and normalized effective mass parameters}
By now it is well known that different particle exchange contributions for $0\nu 2\beta$ decay discussed 
above are also modified by the corresponding nuclear matrix elements which depend upon the chirality of 
the hadronic currents involved \cite{vergados,doi-kotani,bery}.
Including all relevant contributions except those due to doubly charged Higgs exchanges, 
and using eqn.\,({\ref{eq:amp_ll}) - eqn.\,(\ref{eq:amp_lr}}), we express the inverse half-life in terms of effective 
mass parameters with proper normalization factors taking into account the nuclear matrix elements \cite{vergados,doi-kotani,bery} 
leading to the half-life prediction
\begin{eqnarray}
  \left[T_{1/2}^{0\nu}\right]^{-1} &=& G^{0\nu}_{01}\left\{|{\cal M}^{0\nu}_\nu|^2|\eta_\nu|^2 + 
 |{\cal M}^{0\nu}_N|^2|\eta^L_{N_R}|^2 + |{\cal M}^{0\nu}_N|^2|\eta^R_{N_R}|^2  
 \right. \notag \\ & &\left. + |{\cal M}^{0\nu}_\lambda|^2|\eta_\lambda|^2 + 
 |{\cal M}^{0\nu}_\eta|^2|\eta_\eta|^2 \right\} + {\rm interference\ terms}. 
 \label{eq:halflife_simp}
\end{eqnarray}
where the dimensionless particle physics parameters are 
\begin{eqnarray}
\label{eqn:eta1}
& &|\eta_{\nu}| = \left|\frac{\sum_{i}\, \mathcal{V}^{\nu \nu^2}_{ei}\, m_i}{m_e} \right| \nonumber \\
& &|\eta^{R}_{N}| = m_p\,\left(\frac{M_{W_L}}{M_{W_R}}\right)^4\, 
\left|\frac{\mathcal{V}^{NN^2}_{ei}}{M_{N_{i}}} \right|  \nonumber \\
& &|\eta^{L}_{N}| = m_p \, \left|\frac{V^{N\nu}_{ei}}{M_{N_{i}}} + \frac{V^{S\nu}_{ei}}{M_{S_{i}}}\right|  \nonumber \\
& &|\eta_{\lambda}| = \left(\frac{M_{W_L}}{M_{W_R}}\right)^2\, 
\left| U_{ei} \left(\frac{M_D}{M_N}\, \right)_{ei}\, \right| \nonumber \\
& &|\eta_{\eta}| = \tan \zeta_{LR}\, 
\left| U_{ei} \left(\frac{M_D}{M_N} \right)_{ei}\, \right|
\end{eqnarray}
In eqn.\,(\ref{eqn:eta1}), $m_e$ $(m_i)$= mass of electron (light neutrino), and $m_p$ = proton mass. 
In eqn.\,(\ref{eq:halflife_simp}), $G^{0\nu}_{01}$ is the the phase space factor and besides different 
particle parameters, it contains  the nuclear matrix elements due different chiralities of the hadronic 
weak currents such as $\left(\mathcal{M}^{0\nu}_{\nu} \right)$ involving left-left chirality in the standard 
contribution, and due to heavy neutrino exchanges 
$\left(\mathcal{M}^{0\nu}_{\nu} \right)$ involving right-right chirality arising out of heavy neutrino exchange, 
$\left(\mathcal{M}^{0\nu}_{\lambda} \right)$  for the $\lambda-$ diagram, and $\left(\mathcal{M}^{0\nu}_{\eta} 
\right)$ for the $\eta-$ diagram . Explicit numerical values of these nuclear matrix elements discussed 
in ref.\cite{vergados, doi-kotani,bery} are  given in Table.\,\ref{tab:nucl-matrix}.
\begin{table}[h]
 \centering
\vspace{10pt}
 \begin{tabular}{lcccccc}
 \hline \hline
 \\ \multirow{2}{*}{Isotope} & $G^{0\nu}_{01}$ $[10^{-14}\ {\rm yrs}^{-1}]$  & \multirow{2}{*}{${\cal M}^{0\nu}_\nu$} & \multirow{2}{*}{${\cal M}^{0\nu}_N$} & \multirow{2}{*}{${\cal M}^{0\nu}_\lambda$} & \multirow{2}{*}{${\cal M}^{0\nu}_\eta$} \\[1mm] 
 & Refs. \cite{vergados,doi-kotani} & & & \\[0.8mm]
\hline \\
$^{76}$Ge  & 0.686 &2.58--6.64 & 233--412 & 1.75--3.76 & 235--637 \\ 
$^{82}$Se  & 2.95 &2.42--5.92 & 226--408 & 2.54--3.69 & 209--234 \\ 
$^{130}$Te  & 4.13 &2.43--5.04 & 234--384 & 2.85--3.67 & 414--540 \\ 
$^{136}$Xe  & 4.24 &1.57--3.85 & 160--172 & 1.96--2.49 & 370--419 \\ 
\hline \hline
 \end{tabular}
 \caption{Phase space factors and nuclear matrix elements with their allowed ranges as derived in Refs. \cite{vergados,doi-kotani,bery}.}
 \label{tab:nucl-matrix}
\end{table}

In order to arrive at a common normalization factor for all types of contributions, at first we use 
the expression for inverse half-life for $0\nu 2\beta$ decay process due to only light active Majorana neutrinos, 
$\left[T_{1/2}^{0\nu}\right]^{-1} = G^{0\nu}_{01}\left|{\cal M}^{0\nu}_\nu \right|^2|\eta_\nu|^2$. 
Using the numerical values given in Tab\,\ref{tab:nucl-matrix}, we rewrite the inverse half-life 
in terms of effective mass parameter
$$\left[T_{1/2}^{0\nu}\right]^{-1} = G^{0\nu}_{01} \left| \frac{{\cal M}^{0\nu}_\nu}{m_e} \right|^2\, 
|{\large \bf  m}^{\rm ee}_{\nu}|^2 = 1.57 \times 10^{-25}\, \mbox{yrs}^{-1}\, \mbox{eV}^{-2} |{\large 
\bf  m}^{\rm ee}_{\nu}|^2 = \mathcal{K}_{0\nu}\, |{\large \bf  m}^{\rm ee}_{\nu}|^2 $$
where ${\large \bf  m}^{\rm ee}_{\nu} = \sum^{}_{i}  \left(\mathcal{V}^{\nu \nu}_{e\,i}\right)^2\, m_{\nu_i}$.
Then the analytic expression for all relevant contributions to effective mass parameters taking into account the respective 
nuclear matrix elements turns out to be
\begin{eqnarray}
\left[T_{1/2}^{0\nu}\right]^{-1}&=&  \mathcal{K}_{0\nu}\, 
\bigg[ |{\large \bf  m}^{\rm ee}_{\nu}|^2 + |{\large \bf  m}^{\rm ee,R}_{N}|^2
+ |{\large \bf  m}^{\rm ee,L}_{S}|^2 + |{\large \bf  m}^{\rm ee}_{\lambda}|^2 + 
|{\large \bf  m}^{\rm ee}_{\eta}|^2 \bigg] + \cdots
\label{eqn:halflife-eff}
\end{eqnarray}
where the ellipses denote interference terms and all other subdominant contributions. In eqn.\,(\ref{eqn:halflife-eff}), 
the new effective mass parameters are
\begin{eqnarray} 
&&
{\large \bf  m}^{\rm ee,R}_{N} = \sum^{}_{i} \left(\frac{M_{W_L}}{M_{W_R}}\right)^4
\left(\mathcal{V}^{ N N}_{e\,i}\right)^2\, \frac{|p|^2}{m_{N_i}}  \\
&&
{\large \bf  m}^{\rm ee,L}_{S} = \sum^{}_{i} \left(\mathcal{V}^{\nu S}_{e\,i}\right)^2\, \frac{|p|^2}{m_{S_i}} \\
&&
{\large \bf  m}^{\rm ee}_{\lambda} = 10^{-2} \left(\frac{M_{W_L}}{M_{W_R}}\right)^2 
\left| U_{ei} \left(\frac{M_D}{M_N} \cdots \right)_{ei}\right|\, {\bf |p|}\\
& &
{\large \bf  m}^{\rm ee}_{\eta} = \tan \zeta_{LR}\, 
\left| U_{ei} \left(\frac{M_D}{M_N} \cdots \right)_{ei} \right|\, {\bf |p|}
\end{eqnarray}
where $|p|^2 = m_e\,m_p\, \mathcal{M}^{0\nu}_{N}/\mathcal{M}^{0\nu}_{\nu} \simeq (\mbox{200\, MeV})^2$. 
It is to be noted that the suppression factor $10^{-2}$ arises in the $\lambda-$diagram as pointed 
out in refs. \cite{rode-ilc, vergados,doi-kotani,bery}.

\section{Numerical estimation of effective mass parameters}
Using  analytic expression for relevant effective mass parameters given in eqn.\,(3.16)- eqn.\,(3.20) 
and our model parameters discussed in Sec.{\bf 2}, we now estimate the relevant individual contributions numerically.

\subsection{Nearly standard contribution}
In our model the new mixing matrix $\mathcal{N}_{ei} \equiv \mathcal{V}^{\nu \nu}_{e\,i} =
\left(1- \eta \right) U_{\nu}$ contains additional non-unitarity effect due to non-vanishing 
$\eta$ where 
\begin{eqnarray}
\mathcal{N}_{e1} & =&  
                   (1-\eta_{e1})\, U_{11} -\eta_{e2}\, U_{21}
                     -\eta_{e3}\, U_{31} \nonumber \\
\mathcal{N}_{e2} & = &   (1-\eta_{e1})\, U_{12} -\eta_{e2}\, U_{22}
                     -\eta_{e3}\, U_{32} \nonumber \\
\mathcal{N}_{e3} & = &  (1-\eta_{e1})\, U_{13} -\eta_{e2}\, U_{23}
                     -\eta_{e3}\, U_{33} 
\label{eqn:Nee-element}
\end{eqnarray}
We estimate numerical values of ${\cal N}_{ei}$ using all allowed values of $\eta$ discussed in Sec.{\bf 2} 
and also by using $U\equiv U_{\rm PMNS}$.  Then the effective mass parameter for the $W_L-W_L$ mediation with light neutrino exchanges is found to be almost similar to the standard prediction
\bea \label{eq:mee}
|m^{\rm ee}_{\nu}| \simeq \left\{
\baz 
0.004\, \mbox{eV}
& \mbox{ NH,} \\[0.2cm]
0.048\, \mbox{eV}
& \mbox{ IH,} \\[0.2cm]
0.23\, \mbox{eV} 
& \mbox{ QD.} 
\ea \right. 
\eea
This nearly standard 
contribution on effective mass parameter is presented by the dashed-green colored lines of 
Fig.~\ref{fig:znbb_decay1} and Fig.~{\ref{fig:znbb_decay2}} for  NH neutrino masses, but it is 
presented by the dashed-pink colored lines of the same figures for IH neutrino 
masses. 
\begin{figure*}[htb]
\begin{center}
\includegraphics[scale=0.39, angle=-90]{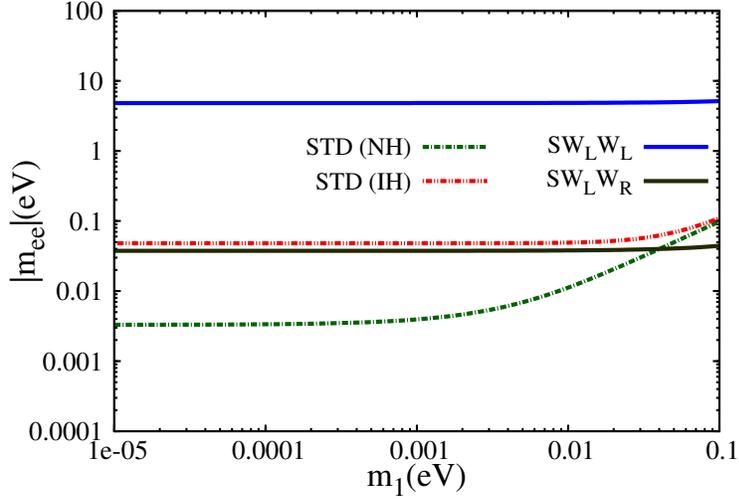}
\caption{Variation of effective mass parameters with lightest neutrino mass. 
        The standard contributions are shown by dashed-green (pink) colored lines  
        for NH (IH) case. The non-standard contribution with $W^-_L-W^-_L$ mediation 
        and sterile neutrino exchanges is shown by the upper blue solid line whereas 
        the one with $W^-_L-W^-_R$ mediation and sterile neutrino exchanges 
        is shown by the lower black solid line.} 
\label{fig:znbb_decay1}
\end{center}
\end{figure*}
In our numerical estimations presented in Fig.\,\ref{fig:znbb_decay1} we have used $M_D$ values including 
RG corrections as given in eq.(\ref{eq:md_with_rge}) but with $M=(50, 200, 1712)$~GeV, $M_N=(1250, 3000, 5000)$~GeV, and 
$m_{\hat{S}}=(2, 13, 532)$~GeV.
 Similarly, in 
Fig.\,\ref{fig:znbb_decay2} we have utilized $M_D$ values including RG corrections from eq.(\ref{eq:md_with_rge}) but with $M=(100, 100, 2151.6)$~GeV, 
$M_N=(5000, 5000, 5000)$~GeV, and $m_{\hat{S}}=(2, 2, 800)$~GeV. 

\subsection{Dominant non-standard contributions} 
Before estimating the non-standard effective mass parameters, we present the mixing matrices numerically. 
As discussed in eq.~(\ref{eqn:Vmix-extended}) 
of Sec.\,{\bf 2}, the mixing matrices $X = M_D\,M^{-1}$, $Y=M\, M^{-1}_N$, $Z=M_D\,M^{-1}_N$, 
and $y=\mu_S\, M^{-1}$ all contribute to non-standard predictions of $0\nu \beta \beta$ 
amplitude in the extended seesaw scheme. 
\begin{figure*}[t]
\begin{center}
\includegraphics[scale=0.39, angle=-90]{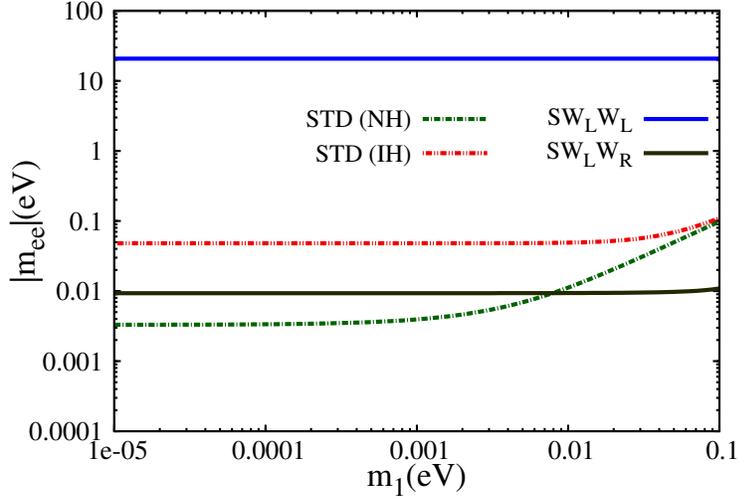}
\caption{Variation of effective mass parameters with lightest neutrino mass. 
        The standard contributions are shown by dashed-green (pink) colored lines 
        for NH (IH) case. The non-standard contribution with $W^-_L-W^-_L$ mediation 
        and sterile neutrino exchanges is shown by the upper blue solid line whereas 
        the one with $W^-_L-W^-_R$ mediation and sterile neutrino exchanges 
        is shown by the lower black solid line.}
\label{fig:znbb_decay2}
\end{center}
\end{figure*}   

Using eqn.(2.13) and the diagonal structures of the RH Majorana neutrino mass matrix $M_N = 
\mbox{diag}(M_{N_1}, M_{N_2}, M_{N_3})$ as well as $N-S$ mixing matrix $M={\rm diag}(M_1, M_2, M_3)$, 
and the Dirac neutrino mass matrix $M_D$ with RG corrections given in eqn.~({\ref{eq:md_with_rge}}), 
we derive the relevant elements of the mixing matrices $\mathcal{N}$, $\mathcal{V}^{\nu N}$, 
$\mathcal{V}^{\nu S}$, $\mathcal{V}^{S \nu}$, $\mathcal{V}^{S S}$, $\mathcal{V}^{S N}$, $\mathcal{V}^{N \nu}$, 
$\mathcal{V}^{N S}$ and $\mathcal{V}^{N N}$ for which one example is
\begin{eqnarray}
& &\hspace*{-0.3cm} \mathcal{N}_{ei}   =    \{0.8135, 0.5597, 0.1278\}\, , \quad  \quad  \quad 
\mathcal{V}^{\nu S}_{ei} =    \{4.5398 \times 10^{-4}, 4.93 \times 10^{-4}, 2.148 \times 10^{-4}\}\, ,\nonumber \\
& &\hspace*{-0.3cm} \mathcal{V}^{\nu N}_{ei} = \{1.8 \times 10^{-5}, 3.3 \times 10^{-5}, 6.7 \times 10^{-5}\}, 
\mathcal{V}^{S \nu}_{ei} = \{3.6 \times 10^{-3}, 3.3 \times 10^{-3}, 6.0 \times 10^{-3}\},\nonumber \\
& &\hspace*{-0.3cm} \mathcal{V}^{S S}_{ei} = \{0.999, 0.0002, 5.0\times 10^{-6}\},\, 
\mathcal{V}^{S N}_{ei} = \{0.04, 0.0, 0.0\},\, 
\mathcal{V}^{NN}_{ei} = \{1.0, 0.0, 0.0 \}\, ,\nonumber \\
& &\hspace*{-0.3cm} \mathcal{V}^{N \nu}_{ei} = \{9.33 \times 10^{-10}, 2.97 \times 10^{-9}, 1.0 \times 10^{-8}\}\, , \quad \quad \quad 
\mathcal{V}^{N S}_{ei} = \{0.04, 0.0, 0.0 \} \,.
\end{eqnarray}
\begin{table}[htb!]
\centering
\begin{tabular}{|c|c|}
\hline 
{\bf C1:} & {\bf C2:} \\
\hline \hline 
\hspace*{-0.6cm} $M=\mbox{diag}\left(50.0, 200.0, 1711\right)$ GeV  & $M=\mbox{diag}\left(100.0, 100.0, 2151.6\right)$ GeV  \\ 
 $M_N= \mbox{diag}\left(1250.0, 3000.0, 5000.0\right)$ GeV & $M_N= \mbox{diag}\left(5000.0, 5000.0, 5000.0\right)$ GeV  \\ 
 $\hat{m}_{S} = \mbox{diag}\left(2.0, 13.0, 532\right)$ GeV & $\hat{m}_{S} = \mbox{diag}\left(2.0, 2.0, 800\right)$ GeV\,.\\
 \hline
\end{tabular}
\caption{Input values of $M$, $M_N$, and $\hat{m}_{S}$ used for estimating 
         effective mass parameters given in Table\,6.}
\label{tab:input-effmass}
\end{table} 
\noindent For evaluating these mixing matrix elements we have taken the input values, $M$, $M_N$, and $\hat{m}_{S}$ 
presented under column ${\bf C1}$ of Table.\,\ref{tab:input-effmass}. These lead to the numerical results for effective mass 
parameter contributing to $0\nu \beta \beta$ decay rate presented under column $\bf C1$ of 
  TABLE.~\ref{tab:effe-mass-para}. Similarly when we use the  $M$, $M_N$, and $\hat{m}_{S}$  values from  column 
  ${\bf C2}$ of Table.\,\ref{tab:input-effmass} we obtain effective mass parameters given in  column ${\bf C2}$ of  TABLE.~\ref{tab:effe-mass-para}.

\begin{table}[htb!]
\centering
\begin{tabular}{|c|c|c|}
\hline
${\bf Effective~ mass~ parameter}$  & {\bf C1~} (\mbox{eV}) & {\bf C2}~ (\mbox{eV})\\ \hline 
$ {\large \bf  m}^{\rm ee}_{\nu}$ & $0.004$ & $0.004$   \\ 
${\large \bf  m}^{\rm ee,R}_{N} $  & $0.0085$  & $0.0085$   \\ 
${\large \bf  m}^{\rm ee,L}_{S}  $  & $20.75$ & $188.48$   \\  
${\large \bf  m}^{\rm ee}_{\lambda,\eta}$  & $\simeq 0.0093$ &$\simeq 0.0274$ \\ 
\hline 
\end{tabular}
\caption{Estimations of effective mass parameter with the allowed model parameters. The results are for the Dirac neutrino 
mass matrix including RG corrections. The input values of mass matrices allowed by 
the current data for different columns are presented in Table\,5.}
\label{tab:effe-mass-para}
\end{table}

The most dominant and new contribution to the effective mass parameters is found to emerge from the 
amplitude $A_S^{LL}$ of eqn.\,(\ref{eq:amp_ll}) due to $W_L^--W_L^-$ mediation and sterile neutrino 
exchanges. This has been shown in Fig.\,\ref{fig:znbb_wl_wl} for various combinations of sterile 
neutrino mass eigenvalues and for $M_D$ values including 
 RG corrections given in eq.\,(\ref{eq:md_with_rge}). 
In Fig.\,\ref{fig:znbb_wl_wl} our estimated values range from 0.2 eV-1.0 eV. Looking to the results 
given in Table.\,\ref{tab:effe-mass-para} and Fig.\,\ref{fig:znbb_decay1}, Fig.\,\ref{fig:znbb_decay2} 
, and Fig.\,\ref{fig:znbb_wl_wl}, it is clear that the actual enhanced rate of $0\nu \beta \beta$ decay in this model depends 
primarily upon the sterile neutrino mass eigenvalues $m_{S_1}$ and $m_{S_2}$. If the decay rate 
corresponds to $|M_{\mathrm{eff}}| \simeq 0.21 - 0.53~ {\rm eV} $ as claimed by the Heidelberg-Moscow 
experiment using $^{76}{Ge}$ \cite{claim}, our new finding is that the light neutrino masses could 
be still of NH or IH pattern, instead of necessarily being of QD pattern,  but with $m_{S_1}\sim 10\, \mbox{GeV}$ and $m_{S_2}\sim 30\, \mbox{GeV}$. Of course the the Dirac neutrino mass matrix having its high scale quark-lepton symmetric origin  also contributes to the magnification of 
the effective mass parameter.
The next dominant contributions coming from the Feynman amplitude 
$A_S^{LR}$ of eqn.~(\ref{eq:amp_lr}) due to $W_L^--W_R^-$ mediation and sterile neutrino exchanges 
with $m^{\rm ee,LR }_{\lambda,\eta} =0.04$~eV (0.01 eV) have been shown in Fig.\,\ref{fig:znbb_decay1} ( Fig.~\ref{fig:znbb_decay2}).
\begin{figure*}[htb!]
\begin{center}
\includegraphics[scale=0.39, angle=-90]{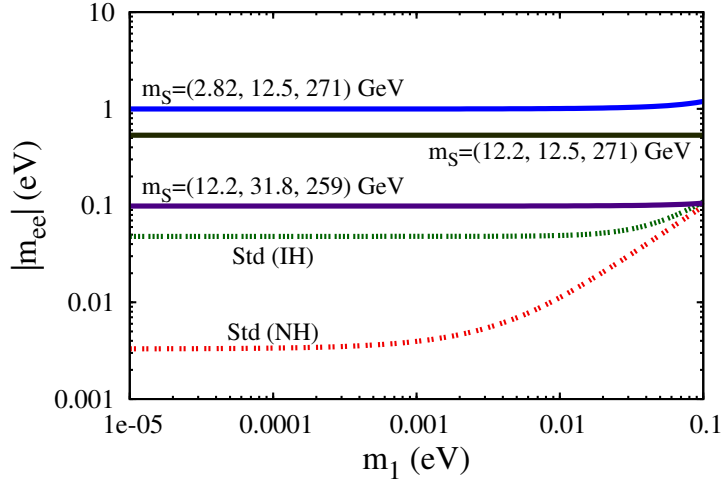}
\caption{Predictions of non-standard contributions to effective mass parameter 
        with $W_L^--W_L^-$ mediation and sterile neutrino exchange for $M=(120, 250, 1664.9)$~GeV 
        (top solid line), $M=(250, 250, 1663.3)$~GeV (middle solid line), and $M=(250, 400, 1626.1)$~GeV 
        (bottom solid line) keeping $M_N=(5, 5, 10)$~TeV fixed and for $M_D$ as in 
        eq.~({\protect\ref{eq:md_with_rge}}). 
        }
\label{fig:znbb_wl_wl}
\end{center}
\end{figure*}

\section{Estimations on lepton flavor violating decays and ${\bf J_{\rm CP}}$}
Besides the neutrinoless double beta decay process, the sterile and heavy neutrinos 
in this model can predominantly mediate different lepton flavor violating decays, 
$\mu \rightarrow e + \gamma$, $\tau \rightarrow e + \gamma$, and $\tau \rightarrow 
\mu + \gamma$. Since $\ell_\alpha \rightarrow \ell_\beta + \gamma$ ($\alpha \neq \beta$) 
is lepton flavor changing process, it is strictly forbidden in the Standard Model when 
$m_\nu=0$ and lepton number is conserved. In our model the underlying lepton-flavor violating
interactions and non-unitarity effects contribute to LFV decays by the mediation of heavy
RH Majorana and sterile Majorana fermions. 
\begin{table}[htb!]
\centering
\begin{tabular}{|l|l|c|c|c|}
\hline\hline 
$M$(GeV)& $M_N$(TeV)& Heavy Mass Eigen Values(GeV)\\ \hline
(9.7, 115.2, 2776.5) & (5, 5, 5) & (0.018, 2.65, 1238, 5000, 5002, 6238)\\ \hline
(100, 100,2151.57) & (5, 5, 5)& (1.99, 2.00, 800.5, 5001, 5002, 5800)\\ \hline
(100, 200, 1702.67) & (5, 5, 5)& (1.99, 8.00, 527.5, 5001, 5007, 5527)\\ \hline
(50, 200,1711) & (1.5, 2, 5)& (1.67, 19.8, 532.2, 1501, 2019, 5532)\\ \hline
(1604.442,1604.442,1604.442) & (5,5,10) &(252.4,461.5,470.6,5471.56,5471.35,10252.4) \\ \hline
\hline 
\end{tabular}
\caption{The Heavy mass eigen values for the matrices of $M$ and $M_N$ which have been used to evaluate branching ratios.}
\label{tab:masses-br}
\end{table}

\subsection{Branching ratio}
Keeping in mind the charged-current interaction in the neutrino mass basis for extended 
seesaw scheme given in eq.\,(\ref{eqn:ccint-flavor}) - eq.\,(\ref{eqn:nustate-mass}), the dominant contributions are mainly through 
the exchange of the sterile and heavy RH neutrinos with branching ratio \cite{lfv-illakovac, mkp-spatra}
\begin{eqnarray}
\label{eq:LFV}
\text{Br}\left(\ell_\alpha \rightarrow \ell_\beta + \gamma \right) =
          \frac{\alpha^3_{\rm w}\, s^2_{\rm w}\, m^5_{\ell_\alpha}}
          {256\,\pi^2\, M^4_{W}\, \Gamma_\alpha} 
           \left|\mathcal{G}^{N}_{\alpha \beta} + \mathcal{G}^{S}_{\alpha \beta}\right|^2 \, ,
\end{eqnarray}
\begin{eqnarray}
&\text{where}~&\mathcal{G}^{N}_{\alpha \beta} =
        \sum_{k} \left(\mathcal{V}^{\nu\, N}\right)_{\alpha\, k}\, 
         \left(\mathcal{V}^{\nu\, N}\right)^*_{\beta\, k} 
         \mathcal{I}\left(\frac{m^2_{N_k}}{M^2_{W_L}}\right) \, ,
         \nonumber \\
& & \mathcal{G}^{S}_{\alpha \beta} = \sum_{j} \left(\mathcal{V}^{\nu\, S}\right)_{\alpha\, j}\, 
         \left(\mathcal{V}^{\nu\, S}\right)^*_{\beta\, j} 
         \mathcal{I}\left(\frac{m^2_{S_j}}{M^2_{W_L}}\right) \, , \nonumber \\
&{\rm and} & \mathcal{I}(x) = -\frac{2 x^3+ 5 x^2-x}{4 (1-x)^3} 
                - \frac{3 x^3 \text{ln}x}{2 (1-x)^4}\, .\nonumber 
\end{eqnarray}
It is clear from the above equation and within the model parameter range, 
$M_N \gg M \gg M_D$, that the first term in eq. (\ref{eq:LFV}) is negligible while second term 
involving the the heavy sterile neutrinos gives dominant contribution which is proportional 
to $\sum_{j} \left(\mathcal{V}^{\nu\, S}\right)_{\alpha\, j}\, \left(\mathcal{V}^{\nu\, S} 
\right)^*_{\beta\, j} \simeq 2 \eta_{\alpha \beta}$. 
\begin{figure}[htb!]
\centering
\includegraphics[scale=0.25, angle=-90]{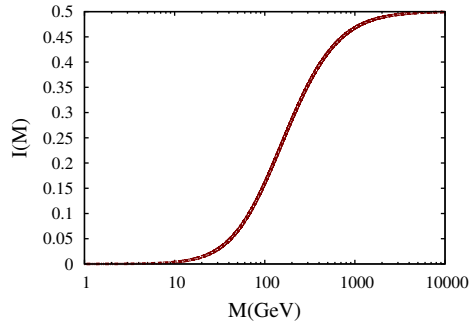} 
\caption{Loop factor vs masses of heavy RH or sterile neutrino.}
\label{fig:log_fact}
\end{figure}

Taking into account the contribution of the non-unitarity matrix, it is clear that out of diagonal 
elements of $M_N$ and $M$, mostly the latter contributes to the branching ratios. 
Also the contribution of loop factor 
for various range of masses allowed in this extended seesaw mechanism is shown in Fig. 
\ref{fig:log_fact}.

\begin{table}[htb!]
\centering
\begin{tabular}{|l|l|c|c|c|}
\hline\hline 
$M$(GeV)& $M_N$(TeV)& Br($\mu\rightarrow e\gamma$) & Br($\tau\rightarrow e \gamma$) & Br($\tau\rightarrow \mu\gamma$)\\ \hline
(50, 200, 1711.8)&(1.5, 2, 5)& $3.05\times 10^{-16}$&$3.11\times 10^{-14}$&$4.36\times 10^{-12}$\\ \hline
(100, 100,2151.57) & (5, 5, 5)& $1.28\times 10^{-16}$&$1.39\times 10^{-14}$&$1.95\times 10^{-12}$\\ \hline
(100, 200, 1702.67) & (5, 5, 5)& $2.85\times 10^{-16}$&$3.1\times 10^{-14}$&$4.3\times 10^{-12}$\\ \hline
(1604.442,1604.442,1604.442) & (5,5,10) &$2.18\times 10^{-16}$ &$2.32\times 10^{-14}$ &$3.25\times 10^{-12}$ \\ \hline
\hline 
\end{tabular}
\caption{The three branching ratios in extended inverse seesaw for different values of $M$ and $M_N$ while $M_D$
is same as in eq.~({\protect\ref{eq:md_with_rge}}).}
\label{tab:braratio}
\end{table}
Using the numerically computed mixing matrix, and using allowed mass scales presented in Table\,\ref{tab:masses-br}, 
our model estimations on branching ratios are given in Table\,\ref{tab:braratio}.
Recent experimental data gives the best limit on these branching ratios for LFV decays 
coming from the MEG collaboration \cite{LFV}. Out of these $\text{Br}\left(\mu \rightarrow e + 
\gamma \right) \leq 1.2 \times 10^{-11}$ \cite{LFV} is almost three orders of magnitude 
stronger than the limit $\text{Br}\left(\tau \rightarrow e + \gamma \right) \leq 3.3 \times 10^{-8}$ or 
$\text{Br}\left(\tau \rightarrow \mu + \gamma \right) \leq 4.4 \times 10^{-8}$
at 90$\%$ C.L. However, projected reach of future sensitivities of ongoing searches 
are $\text{Br}\left(\tau \rightarrow e + \gamma \right) \leq 10^{-9}, ~ \text{Br}\left(\tau 
\rightarrow \mu + \gamma \right) \leq 10^{-9} $, and $\text{Br}\left(\mu \rightarrow e + 
\gamma \right) \leq 10^{-18}$ \cite{LFV} which might play crucial role in verifying or falsifying 
the discussed scenario.
\subsection{CP-violation due to non-unitarity}
There are attempts taken in long baseline experiments \cite{DAYA-BAY} with accelerator neutrinos 
$\nu_{\mu}$ and anti-neutrons $\bar{\nu}_{\mu}$ to search for CP violating effects in neutrino oscillations. 
In the usual notation, the standard contribution to these effects is determined by the rephasing invariant 
$J_{\rm CP}$ associated with the Dirac phase $\delta_{CP}$ and matrix elements of the PMNS matrix 
$$J_{\rm CP}\equiv \text{Im}\left(U_{\alpha\, i}U_{\beta\, j} U^*_{\alpha\,j} U^*_{\beta\, j}\right) 
= \cos \theta_{12}\, \cos^2 \theta_{13}\, \cos \theta_{23}\, 
                      \sin \theta_{12}\, \sin \theta_{13}\, \sin \theta_{23}\, \sin \delta_{\rm CP}.$$ 
In this extended seesaw mechanism, the leptonic CP-violation can be written as
\begin{eqnarray}
\mathcal{J}^{ij}_{\alpha \beta} =  \text{Im}\left(\mathcal{N}_{\alpha\, i} 
\mathcal{N}_{\beta\, j} \mathcal{N}^*_{\alpha\,j} \mathcal{N}^*_{\beta\, j}\right)
\simeq J_{\rm CP} + \Delta J^{ij}_{\alpha \beta}\, 
\end{eqnarray}
where \cite{Bdev-non, ramlal, antbnd, non-unit}
\begin{eqnarray}
\Delta J^{ij}_{\alpha \beta} = - \sum_{\rho=e, \mu, \tau} & &\text{Im} \bigg[  
                      \eta_{\alpha \rho}\, U_{\rho i}\, U_{\beta j}\, U^*_{\alpha j}\, U^*_{\beta i}  
                    + \eta_{\beta \rho}\, U_{\alpha i}\, U_{\rho j}\, U^*_{\alpha j}\, U^*_{\beta i} 
\nonumber \\  &&     + \eta^*_{\alpha \rho}\, U_{\alpha i}\, U_{\beta j}\, U^*_{\rho j}\, U^*_{\beta j} 
                    + \eta^*_{\beta \rho}\, U_{\alpha i}\, U_{\beta j}\, U^*_{\alpha j}\, U^*_{\rho j} \bigg]\, .
\end{eqnarray}
The extra contribution arises because of the non-unitarity mixing matrix which depends on both $M_D$ and 
$M$. Thus the new contribution to CP-violation is larger for larger $M_D$ which is generated with 
quark-lepton symmetry and for smaller $M$ while safeguarding the constraint $M_N \gg M > M_D, \mu_S$. 
It is noteworthy that in our model even if the leptonic Dirac phase $\delta_{\rm CP} \simeq 0, \pi, 2 \pi$, and/or 
$\sin \theta_{13} \to 0$, there is substantial contribution to CP-violation which might arise out of 
the imaginary parts of the non-unitarity matrix elements $\eta_{\alpha \beta}$.

\begin{figure}[htb!]
\begin{minipage}[t]{0.48\textwidth}
\hspace{-0.4cm}
\begin{center}
\includegraphics[scale=.28, angle=-90]{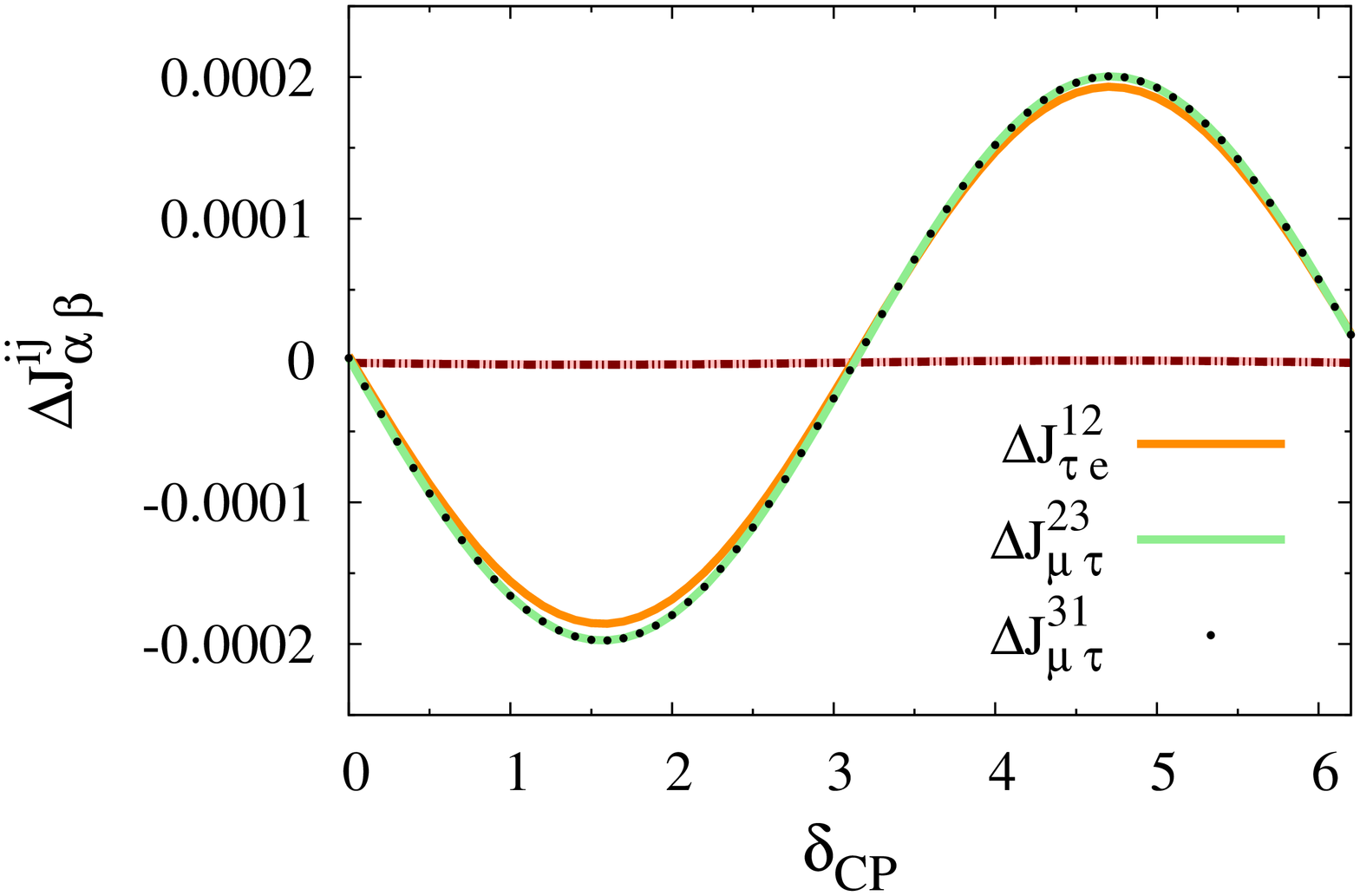}
\end{center}
 \end{minipage}
 \hfill
 \begin{minipage}[t]{0.48\textwidth}
 \hspace{-0.4cm}
 \begin{center}
 \includegraphics[scale=.28, angle=-90]{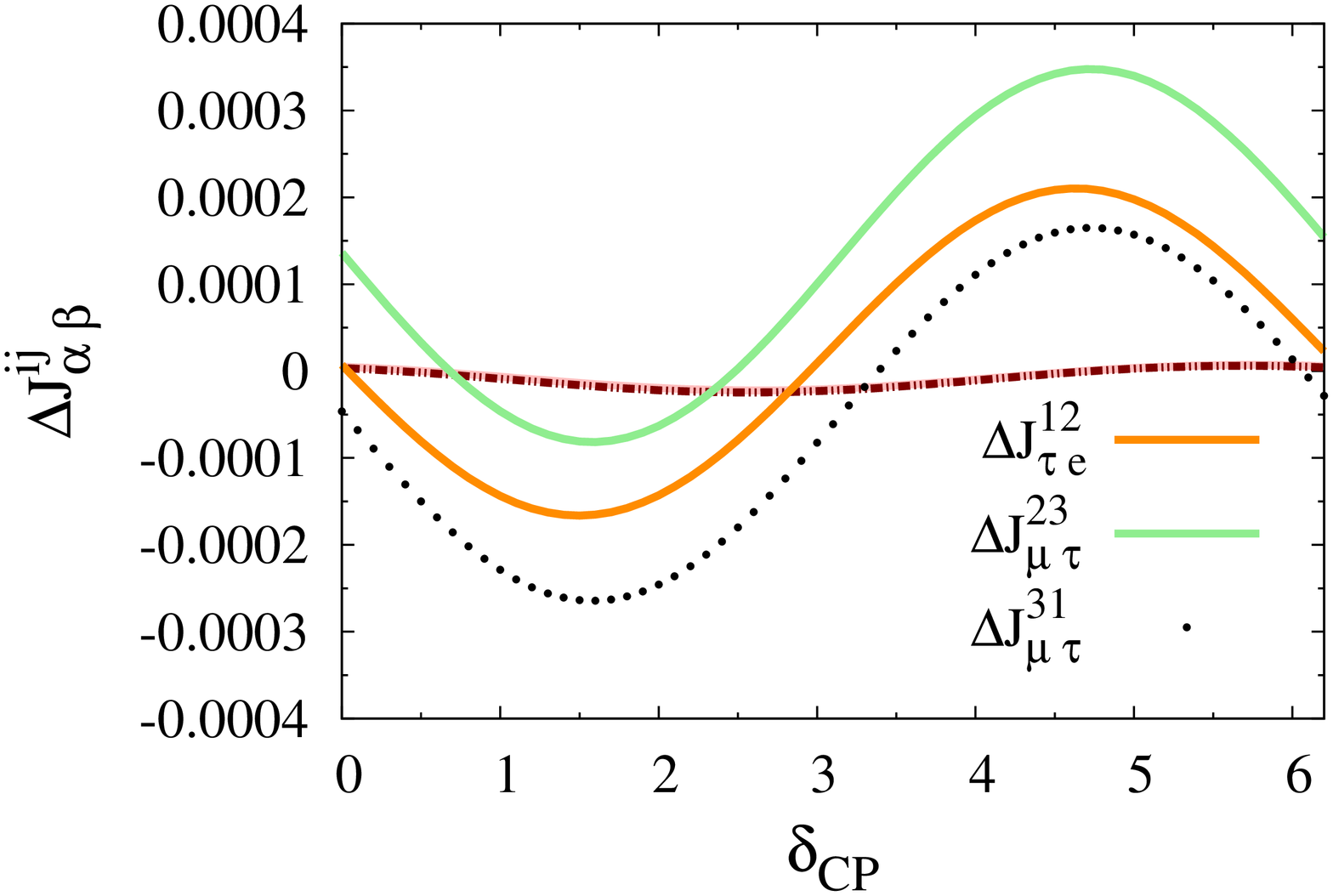}
 \end{center}
 \end{minipage}
 \caption{CP-violation for the full allowed range of leptonic Dirac phase $\delta_{CP}$. 
         The left-panel corresponds to degenerate values of $M$ with $M_1=M_2=M_3\simeq 1604.442$ 
         GeV, and the right panel is due to non-degenerate $M$ with $M_1=9.7$ GeV, $M_2=115.2$ 
         GeV, and $M_3=2776.5$ GeV. }
\label{fig:jcp}
 \end{figure}

\begin{table}[htb]
\centering
\begin{tabular}{|c|c|c|c|c|c|}
\hline\hline
M&$\Delta\mathcal{J}^{12}_{e\mu}$& $\Delta\mathcal{J}^{23}_{e\mu}$
&$\Delta\mathcal{J}^{23}_{\mu\tau}$&$\Delta\mathcal{J}^{31}_{\mu\tau}$&$\Delta\mathcal{J}^{12}_{\tau e}$ \\ \hline
  (a) &$-2.0\times 10^{-6}$ &$-2.3\times 10^{-6}$ &$-1.2\times 10^{-4}$
&$-1.2\times 10^{-4}$&$-1.1\times 10^{-4}$ \\ 
 (b) &$-2.7\times 10^{-6}$ &$-3.2\times 10^{-6}$&$-1.2\times 10^{-4}$ &$-1.2\times 10^{-4}$&$-1.1\times 10^{-4}$\\ 
 (c) &$-2.1\times 10^{-5}$ &$-2.4\times 10^{-5}$ &$1.1\times 10^{-7}$&$-1.8\times 10^{-4}$&$-7.9\times 10^{-5}$ \\ \hline
\hline
\end{tabular}
\caption{The CP-violating effects for (a) degenerate masses $M=(1604.442,1604.442,1604.442)$~GeV, (b) partially degenerate masses  
$M=(100, 100,2151.57)$~GeV and (c) non degenerate masses $M=(9.7, 115.2, 2776.5)$~GeV, while $M_D$ is same as in 
eq.~({\protect\ref{eq:md_with_rge}}).}
\label{tab:angles}
\end{table}

Our estimations using RG corrected Dirac neutrino mass matrix and both degenerate 
and non-degenerate matrix $M$ are shown in the left-panel and right-panel of Fig. 
\ref{fig:jcp}. If the leptonic Dirac phase $\delta_{\rm CP} \neq 0, \pi, 2 \pi$, 
significant CP-violation up to $|\Delta\, J|_{\rm max} \simeq 1.5 \times 10^{-4}$ 
is found to occur for degenerate $M$, but when $M$ is non-degenerate we obtain 
$|\Delta\, J|_{\rm max} \simeq (2-4) \times 10^{-4}$. Also even if $\delta_{\rm CP} 
\simeq 0, \pi, 2 \pi$, non-vanishing CP-violation to the extent of $|\Delta\, J| 
\simeq (1-2) \times 10^{-4}$ is noted to emerge for non-degenerate $M$. These results may be compared with 
CP-violation in the quark sector where $\mathcal{J}_{\rm CKM} \simeq 3.05^{+0.19}_{-0.20} \times 10^{-5}$ 
\cite{PDG} which is nearly one order lower than the leptonic case. 
The horizontal lines in Fig.\,\ref{fig:jcp} represent absence of non-unitarity effects on 
CP-violation. 
\section{Implementation in SO(10)}
Our main goal in this section is to examine whether the TeV scale LR gauge model that 
has been shown to give rise to dominant contribution to $0\nu\beta\beta$ decay and lepton 
flavor violation in Sec.{\bf 2} - Sec.{\bf 5} can emerge from a non-SUSY $SO(10)$ grand unified theory. 
Although the search for low mass $W_R^\pm$ bosons in non-SUSY GUTs has been attempted initially 
without \cite{rizzo} precision CERN-LEP data on $\alpha_S(M_Z)$ and $\sin^2\theta_W(M_Z)$ 
\cite{PDG}, there are more recent results on physically appealing intermediate scales 
\cite{barto,Dparity22}. But the analyses in non-SUSY cases where the $B-L$ breaking scale 
synonymous to $W_R$ gauge boson mass much lower than $10^{10}$~GeV are ruled out because 
of the associated large contributions to light neutrino masses via type-I seesaw mechanism. 
In view of the rich phenomenological consequences of the extended seesaw mechanism that 
evades the discordance between dominant $0\nu \beta \beta$ decay and small neutrino mass 
predictions as discussed in Sec.{\bf 2} - Sec.{\bf 5}, we explore the possibility of such low scale 
LR gauge theory in the minimally extended $SO(10)$ grand unification model. 
\subsection{Symmetry breaking chain}
We consider the symmetry breaking chain discussed in Ref. \cite{Dparity22}. Although this model, 
as such, is ruled out because of the TeV scale canonical seesaw that operates to give large neutrino 
masses in contravention of the oscillation data, here we modify this model by including the additional 
doublets $(\chi_L, \chi_R)\subset 16_H$ of $SO(10)$ and extending the minimal fermion content in $\{16\}_F$ 
with the addition of one $SO(10)$ singlet neutral fermion per generation in order to implement the extended 
seesaw mechanism
\begin{eqnarray}
SO(10)
  & &\mathop{\longrightarrow}^{M_U}_{\{54\}} SU(2)_L \times SU(2)_{R} \times SU(4)_C \times D 
                \quad \quad \left[\mathcal{G}_{224D}, \, \, (g_{2L} = g_{2R})  \right]\nonumber \\ 
& &\mathop{\longrightarrow}^{M_P}_{\{210\}} SU(2)_L \times SU(2)_{R} \times SU(4)_C
                 \quad \quad \left[\mathcal{G}_{224}, \, \, (g_{2L} \neq g_{2R})  \right]\nonumber \\ 
& &\mathop{\longrightarrow}^{M_C}_{\{210\}} SU(2)_L \times SU(2)_{R} \times U(1)_{B-L} \times SU(3)_C
                 \quad \quad \left[\mathcal{G}_{2213} \, \right]\nonumber \\ 
& &\mathop{\longrightarrow}^{M^+_R}_{\{210\}} SU(2)_L \times U(1)_{R} \times U(1)_{B-L} \times SU(3)_C
                 \quad \quad \left[\mathcal{G}_{2113} \, \right]\nonumber \\ 
& &\hspace*{-0.2cm} \mathop{\longrightarrow}^{M^0_R}_{\{126+16 \}} SU(2)_L \times U(1)_{Y} \times SU(3)_C
                \quad \quad \left[\mathcal{G}_{\rm SM} \, \right] \nonumber \\
& &\mathop{\longrightarrow}^{M_Z}_{\{10\}}~U(1)_{\rm em}\times SU(3)_C.  
\label{chain}              
\end{eqnarray}
It was found that the $G_{224}$-singlets in $\{54\}_H$ and $\{210\}_H$ of $SO(10)$ are D-parity 
even and odd, respectively. Also it was noted that the neutral components of the $G_{224}$ multiplet 
$\{1, 1, 15\}$ contained in $\{210\}_H$ and $\{45\}_H$ of $SO(10)$ have D-parity even and odd, 
respectively. In the first step, VEV is assigned along the $\langle(1, 1, 1)\rangle\subset\{54\}_H$ 
which has even D-Parity to guarantee the LR symmetric Pati-Salam group to survive while at the second 
step D-parity is broken by assigning $\langle(1, 1, 1) \rangle \subset\{210\}_H$ to obtain asymmetric 
$G_{224}$ with $g_{2L}\neq g_{2R}$. The spontaneous breaking $G_{224}\rightarrow G_{2213}$ is achieved 
by the VEV $\langle(1, 1, 15)^0_H\rangle\subset\{210\}_H$. The symmetry breaking $G_{2213} \rightarrow 
G_{2113}$ is implemented by assigning $O(M_R^+)$  VEV to the neutral component of the sub-multiplet 
$\langle(1, 3, 15)^0_H\rangle \subset \{210\}_H$, and the breaking $U(1)_R\times U(1)_{B-L}\rightarrow 
U(1)_Y$ is achieved by $\langle\Delta^0_R\rangle \subset\{126\}_H$ while the VEV $\langle\chi_R^0\rangle 
\subset \{16\}_H$ provides the $N$-$S$ mixing. As usual, the breaking of SM to low energy symmetry 
$U(1)_{em}\times  SU(3)_C$ is carried out by the SM doublet contained in the bidoublet $\Phi\subset 
\{10\}_H$.
\begin{table}[b]
\centering
\begin{tabular}{|l|l|l|c|c|c|}
\hline
$M_R^0$~(TeV)&$M^+_R$~(TeV)&$M_C$~(TeV)&$M_P$~(GeV)&$M_G$~(GeV)&$\alpha_G$\\
\hline
5&10&$10^3$&$10^{14.2}$&$10^{17.64}$&0.03884\\
\hline
5&10&$10^{3.5}$&$10^{14.42}$&$10^{17.61}$&0.03675\\
\hline
5&20&$10^{3}$&$10^{14.08}$&$10^{17.54}$&0.03915\\
\hline
5&10&$100$&$10^{13.72}$&$10^{17.67}$&0.0443\\
\hline
5&20&$500$&$10^{13.93}$&$10^{17.55}$&0.0406\\
\hline
\end{tabular}
\caption{Predictions of allowed mass scales and the GUT couplings in the $SO(10)$ symmetry breaking chain with low-mass $W_R^{\pm}, Z_R$ bosons.} 
\label{tab:gauge_gut}
\end{table}
\subsection{Gauge coupling unification}
We have evaluated the one-loop and two-loop coefficients of $\beta$-functions of renormalization 
group equations for the gauge couplings \cite{jones} 
\begin{equation}
\mu\,\frac{\partial g_{i}}{\partial \mu}=\frac{a_i}{16 \pi^2} g^{3}_{i} + 
\frac{1}{16 \pi^2}\, \sum_{j} b_{ij} g^3_{i} g^2_{j}\, ,
\end{equation}
and they are given in Table. {\ref{tab:beta_coeff} of appendix.

The Higgs spectrum used in different ranges of mass scales under respective gauge symmetries (G) are
\begin{eqnarray}
& &\hspace*{-1.0cm} {\bf \mbox{(i)}\, \mu=M_Z - M_R^0 }:  G={\rm SM} = G_{213}, 
             \hspace*{0.2cm} \Phi (2,1/2,1)\, ; \nonumber \\
& &\hspace*{-1.0cm} {\bf \mbox{(ii)}\, \mu=M_R^0 - M_R^+}: G= G_{2113}, 
             \hspace*{1.1cm} \Phi_1 (2,1/2,0,1) \oplus \Phi_2 (2,-1/2,0,1) \oplus \nonumber \\ 
& &          \hspace*{5.6cm} \chi_R (1,1/2,-1,1) \oplus \Delta_R (1,1,-2,1) \, ; \nonumber \\ 
& &\hspace*{-1.0cm} {\bf \mbox{(iii)}\, \mu=M_R^+ - M_C}: G= G_{2213},
             \hspace*{1.1cm} \Phi_1 (2,2,0,1) \oplus \Phi_2 (2,2,0,1) \oplus \nonumber \\ 
& &          \hspace*{5.6cm} \chi_R (1,2,-1,1)\oplus \Delta_R (1,3,-2,1)\oplus \Sigma_R(1,3,0,1)\, ; \nonumber \\ 
& &\hspace*{-1.0cm} {\bf \mbox{(iv)}\, \mu=M_C - M_P}: G= G_{224},
             \hspace*{1.3cm} \Phi_1 (2,2,1) \oplus \Phi_2 (2,2,1) \oplus \nonumber \\ 
& &          \hspace*{5.6cm} \chi_R (1,2, {\bar 4}) \oplus \Delta_R (1,3,{\bar {10}}) \oplus \Sigma_R (1,3,15)\, ; \nonumber \\ 
& &\hspace*{-1.0cm} {\bf \mbox{(v)}\, \mu=M_P - M_U}: G= G_{224D},
             \hspace*{1.1cm} \Phi_1 (2,2,1) \oplus \Phi_2 (2,2,1) \oplus \nonumber \\ 
& &          \hspace*{5.6cm} \chi_L (2,1,4) \oplus \chi_R (1,2, {\bar 4}) \oplus \nonumber \\ 
& &          \hspace*{5.6cm} \Delta_L (3,1,10) \oplus \Delta_R (1,3,{\bar {10}}) \oplus \nonumber \\ 
& &          \hspace*{5.6cm} \Sigma_L(3,1,15)\oplus \Sigma_R (1,3,15)\,.
\end{eqnarray}                  
Recently bounds on the masses of the charged and neutral components of the second Higgs 
doublet in the left-right symmetric model has been estimated to be $O(20)$ TeV \cite{bound-2higgs}. 
While searching for possible mass scales we have used the second Higgs doublet $\Phi_2$ 
only for $\mu 
\geq 10$ TeV. 

We have used extended survival hypothesis in implementing spontaneous symmetry breaking of $SO(10)$ 
and intermediate gauge symmetries leading to the SM gauge theory \cite{aguilaibanez,rnmgoran}.
In addition to D-Parity breaking models \cite{Dparity11,Dparity22}, the importance of the Higgs 
representation ${210}_H$ has been emphasized in the construction of a minimal supersymmetric $SO(10)$
GUT model \cite{minso10}. But the present non-SUSY $SO(10)$ symmetry breaking chain 
shows a departure in that the $G_{224D}$ symmetry essentially required  at the highest intermediate 
scale has unbroken D-Parity which is possible by breaking the GUT symmetry through the Higgs representation 
${54}_H\subset SO(10)$ that acquires GUT-scale VEV in the direction of its D-parity even $G_{224D}$-singlet. The importance 
of this $G_{224D}$ symmetry in stabilizing the values of $M_P$ and $\sin^2\theta_W (M_Z)$ against GUT-Planck 
scale threshold effects has been discussed in ref.\,\cite{theorem} and Sec.\,6.4 below.

Using precision CERN-LEP data \cite{PDG} $\alpha_S(M_Z)=0.1184,\, \sin^2\theta_W(M_Z)=0.2311$ 
and $\alpha^{-1}(M_Z)=127.9$, different allowed solutions presented in Table. \ref{tab:gauge_gut}. 
One set of solutions corresponding to low mass $W_R^\pm$ and $Z_R$ gauge bosons is
\begin{figure}
\begin{center}
\includegraphics[width=11cm, height=13cm, angle=-90]{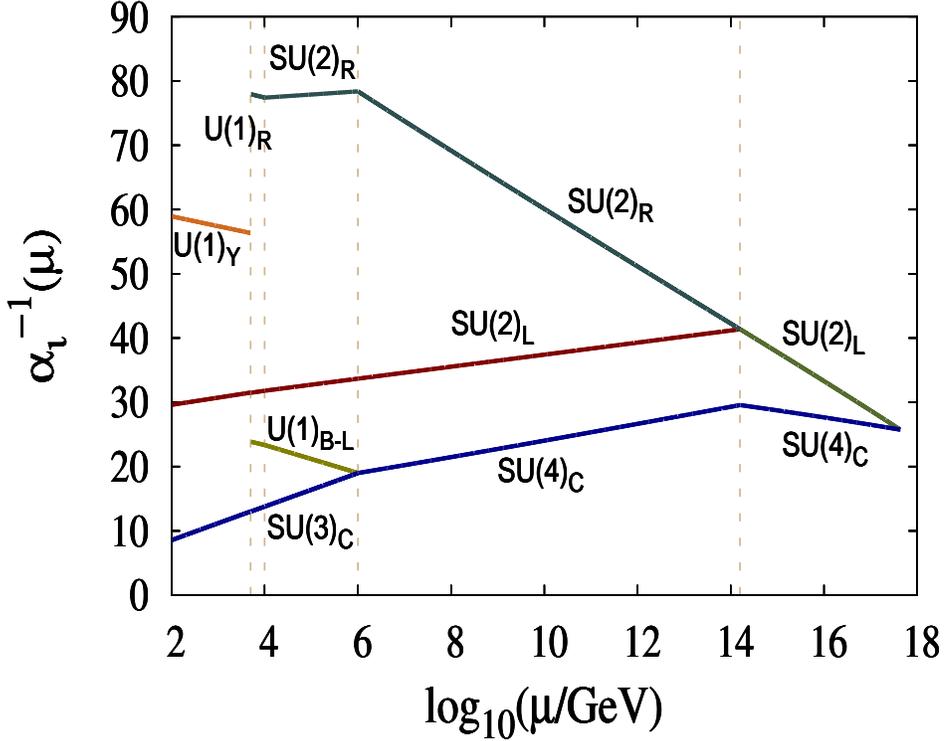}
\end{center} 
\caption{Two loop gauge coupling unification in the $SO(10)$ symmetry breaking chain described in the text. These results are also valid with $G_{224D}$ symmetry near GUT-Planck scale.}
\label{fig:gauge_unif}
\end{figure}
\bea 
M_R^0= 3-5~{\rm TeV},\, M_R^+= 10~{\rm TeV},\, M_C = 10^{2}\, \mbox{TeV}- 10^{3}\, \mbox{TeV}\, ,\nn \\
M_P\simeq 10^{14.17}~{\rm GeV\,\, and\,\,} M_{\rm G}\simeq 10^{17.8}~{\rm GeV}.
\eea
For these mass scales the emerging pattern of gauge coupling unification is shown in Fig.\,\ref{fig:gauge_unif} with GUT fine structure constant $\alpha_G = 0.0388$.

\subsection{Physical significance of mass scales}
The presence of $G_{224D}$ symmetry above the highest intermediate scale plays a crucial role
in lowering down the values of $M_R^+$ while achieving high scale gauge coupling unification. 
With the gauge couplings allowed in the region $\mu\simeq \mbox{3\, TeV\,-\, 10\, TeV}$ 
in the grand unified scenario with $g_{\rm B-L} \simeq 0.725$, $g_{\rm 2R} \simeq 0.4$, 
we have estimated the predicted $W_R$ and $Z_R$ masses to be $M_{W_R} \simeq 4$ TeV, 
$M_{Z_R} \simeq (2.3-3.6)$ TeV for the allowed mass scales $M^0_{R} \simeq (3-5)$ TeV, 
and $M^+_{R} \simeq 10$ TeV of Table.\,\ref{tab:gauge_gut}. 
These low mass $W_R$ and $Z_R$ bosons have interesting RH current effects at low energies 
including $K_L$-$K_S$ mass difference and dominant $0\nu\beta\beta$ rates as discussed in 
Sec.\,{\bf 3}- Sec.\,{\bf 5}. The predicted low mass $W_R^\pm$ and $Z_R$ bosons are also expected to be 
testified at the LHC and future accelerators for which the current bounds are $M_{W_R} \geq 
2.5$ TeV \cite{WR-limit,bound:wr} and $M_{Z_R} \geq 1.162$ TeV \cite{bound:mzr1,bound:mzr2}. The predicted mass scale $M_C\sim \left(10^{5}- 
10^{6} \right)$ ~GeV leads to experimentally verifiable branching ratios for rare kaon decay with Br$(K_L^0\rightarrow 
\mu\ovl{e})\simeq \left(10^{-9}-10^{-11} \right)$ \cite{rarek} via leptoquark gauge boson mediation \cite{pp}. 
Because of the presence of $\mathcal{G}_{224}$ symmetry for $\mu \geq M_C \left(10^{5}-10^{6} 
\right)$ GeV, all the components of diquark Higgs scalars in $\Delta_R \left(3, 1, \bar{10} 
\right)$ mediating $n-\bar{n}$ and $H-\bar{H}$ oscillations also acquire masses at that scale 
whereas the dilepton Higgs scalar carrying $B-L=-2$ is at the $\simeq 1$ TeV scale. This gives rise 
to observable $n-\bar{n}$ oscillation with mixing time $\tau_{n-\bar{n}} \simeq \left(10^{8} - 
10^{11}\right)$ secs \cite{nnbar,nnbar-moh}. However because of the large value of the GUT scale $M_G \simeq 10^{18}$ 
GeV, which is close to the Planck scale, the predicted proton life time for $p \rightarrow e^+\, 
\pi^0$ is large, i.e. $\tau_{\rm p} \geq 10^{40}$ yrs which is beyond the accessible range of ongoing 
search experiments that have set the lower limit $\left(\tau_{\rm p} \right)\big|_{\rm expt.} 
\geq 1.1 \times 10^{34}$ yrs \cite{nishino}. 
\subsection{Importance of $\mathcal{G}_{224D}$ intermediate symmetry}
Near Planck scale unification of this model exposes an interesting possibility that grand 
unification can be also achieved by the Pati-Salam symmetry $\mathcal{G}_{224D}$ even without
the help of the GUT-gauge group SO(10) since, above this scale, gravity effects are expected to 
take over \cite{g224GUT}.

The most interesting role of $G_{224D}$ gauge symmetry at the highest intermediate scale has been 
pointed out in Ref. \cite{theorem}. Normally super-heavy Higgs scalars contained in larger representations 
like $\{210\}_H$ and $\{126\}_H$ introduce uncertainties into GUT predictions of $\sin^2\theta_W 
(M_Z)$ on which CERN-LEP data and others have precise experimental results. But the presence of $G_{224D}$ 
at the highest scale achieves the most desired objective that the GUT scale corrections to $\sin^2 
\theta_{W}(M_Z)$ vanish due to such sources as super-heavy particles or higher dimensional operators 
signifying the effect of gravity.  
\subsection{Determination of Dirac neutrino mass matrix}
It is well known that within Pati-Salam gauge symmetry $\mathcal{G}_{224D}$, the presence of $SU(4)_C$ 
unifies quarks and leptons treating the latter as fourth color and this relates the up-quark mass matrix ($M^0_u$) 
to the Dirac neutrino mass matrix $M^0_D$ at the unification scale. Such relations are also valid in $SO(10)$ 
at the GUT scale since $\mathcal{G}_{224D}$ is its maximal subgroup. Over the recent years it has been shown 
that in a large class of $SO(10)$ model the fermion mass fits at the GUT scale gives $M^0_D \sim {\cal O}(M^0_u)$ 
\cite{goh,Bdev-non,ramlal}. Since the predictions of lepton number and lepton flavor violations carried out in this 
work are sensitive to the Dirac neutrino mass matrix, it is important to derive $M_D$ at the TeV scale given 
in eq.\,(\ref{eq:md_with_rge}). This question has been 
answered in non-SUSY $SO(10)$ \cite{ramlal} and SUSY $SO(10)$ \cite{Bdev-non} while utilizing renormalization 
group running of fermion masses analogous to ref. \cite{dp} and using their low energy data but in the presence 
of intermediate symmetries $G_{2113}$, $G_{2213}$, and $G_{2213D}$. In this analysis we will also use additional 
RGEs for Yukawa coupling and fermion masses in the presence of $G_{224}$ and $G_{224D}$ symmetries operating 
between $M_C \simeq 10^{5}$ GeV to $M_{\rm GUT} \simeq 10^{17.5}$ GeV \cite{pu}.

The determination of the Dirac neutrino mass matrix $M_D(M_{R^0})$ at the TeV seesaw scale is done in three 
steps \cite{ramlal}: ({\bf A.}) Extrapolation of masses to the GUT-scale using low-energy data on fermion masses 
and CKM mixings through corresponding RGEs in the bottom-up approach, ({\bf B.}) Fitting the fermion masses at the GUT scale and determination 
of $M_D(M_{GUT})$, ({\bf C.}) Determination of  $M_D(M_{R^0})$ by top-down approach.
\begin{center}
\subsubsection*{A.\,Extrapolation of fermion masses to the GUT scale}
\end{center}

\noindent At first RGEs for Yukawa coupling matrices and fermion mass matrices are set up from which RGEs 
for mass eigen values and CKM mixings are derived in the presence of $G_{2113}$, $G_{2213}$, $G_{224}$, 
and $G_{224D}$ symmetries.

Denoting $\Phi_{1,2}$ as the corresponding bidoublets under $G_{2213}$ their VEVs are taken as

\bea
\langle \Phi_1 \rangle &=& \begin{pmatrix}v_u&0\\0&0\end{pmatrix},\nonumber\\
\langle \Phi_2 \rangle &=& \begin{pmatrix}0&0\\0&v_d\end{pmatrix}.\label{vevdef}  
\eea

For mass scales $\mu \ll M_G$, ignoring the contribution of the superheavy bidoublet in ${126}_H$, 
the bidoublet $\Phi_1 \subset {10}_{H_1}$ is assumed to give dominant contribution to up quark and 
Dirac neutrino masses $M_u$ and $M_D$ whereas $\Phi_2 \subset {10}_{H_2}$  is used to generate 
masses for down quarks and charged leptons, $M_d$ and $M_\ell$
\bea
M_u &=& Y_u\,v_u,~~M_D=Y_{\nu}\,v_u,~~M_d=Y_d\,v_d,\nonumber\\
M_e &=& Y_e\,v_d, \quad M_R=y_{\chi}\, v_{\chi},\label{defmass} 
\eea  
At $\mu= M_Z$ we use the input values of running masses and quark mixings as in ref.\cite{dp}
\bea
m_e     &=& 0.48684727 \pm 0.00000014\, \mbox{MeV},\, m_{\mu}=102.75138\pm 0.00033\,\mbox{MeV},\nn\\ 
m_{\tau}&=& 1.74669^{+0.00030}_{-0.00027}\,\mbox{GeV},\, m_d=4.69^{+0.60}_{-0.0.66}\,\mbox{MeV},\nn\\  
m_s     &=& 93.4^{+11.8}_{-13.0}\,\mbox{MeV},\, m_b=3.00\pm 0.11\,\mbox{GeV},\nn\\ 
m_u     &=& 2.33^{+0.42}_{-0.45}\,\mbox{MeV},\, m_c=677^{+56}_{-51}\, \mbox{MeV},\nn\\  
m_t     &=& 181\pm 1.3\, \mbox{GeV},\nn\\
\theta^q_{12} &=& {13.04}^{\circ}\pm {0.05}^{\circ},\, \theta^q_{13}={0.201}^{\circ}\pm{0.201}^{\circ},\nn\\   
\theta^q_{23} &=& {2.38}^{\circ}\pm {0.06}^{\circ},\label{zdata}
\eea    
with the CKM Dirac phase  $\delta^q=1.20\pm 0.08$. This results in the CKM matrix at $\mu= M_Z$
\bea
V_{\rm CKM} &=& \begin{pmatrix} 0.9742&0.2256&0.0013-0.0033i\\
-0.2255+0.0001i&0.9734&0.04155\\
0.0081-0.0032i&-0.0407-0.0007i&0.9991\end{pmatrix}.\label{vckz} 
\eea
We use RGEs of the standard model \cite{dp} to evolve all charged fermion mases and CKM mixings 
from $\mu=M_Z$ to $M_R^0\simeq 10$ TeV. With two Higgs doublets at $\mu > 10$ TeV consistent with the 
current experimental lower bound on the second Higgs doublet \cite{bound-2higgs}, we use the starting 
value of $\tan\beta=v_u/v_d=10$  and evolve the masses up to $\mu = M_C$ using RGEs 
derived in the presence of non-SUSY $SO(10)$ and intermediate symmetries $G_{2113}$ and $G_{2213}$ 
\cite{ramlal} with two Higgs bidoublets. For $\mu \ge M_C$, we use the fermion mass RGEs in the presence of $G_{224}$ and 
$G_{224D}$ \cite{pu} modified including the corresponding RGEs of $v_u$ and $v_d$. The fermion mass 
eigen values $m_i$ and the $V_{CKM}$ at the GUT scale turn out to be\\

\noindent{\large {\mbox{At}\, $\mu = {\rm M_{GUT}}$} scale:}
\begin{eqnarray}
m^0_e &=&0.2168~{\rm MeV}, m^0_{\mu}={\rm 38.846}~{\rm MeV}, m^0_{\tau}=0.9620~
{\rm GeV},\nonumber\\
m^0_d&=&1.163~{\rm MeV},m^0_s=23.352~{\rm MeV}, m^0_b=1.0256~{\rm GeV},\nonumber\\ 
m^0_u&=&1.301~{\rm MeV}, m^0_c =0.1686~{\rm GeV}, m^0_t = 51.504~{\rm GeV}\, , 
\label{eqn:eigenu}
\end{eqnarray}
\begin{eqnarray}
{\small V^0_{\rm CKM}(M_{\rm GUT})}= {\small \begin{pmatrix} 0.9764& 0.2160& -0.00169-0.00356i\\
-0.2159-0.0001i& 0.9759-0.00002i& 0.0310\\
0.00835-0.00348i& -0.02994-0.00077i& 0.9995\end{pmatrix}},
\label{eqn:vckmu} 
\end{eqnarray}
where, in deriving eqn.\,(\ref{eqn:eigenu}), we have used ``run and diagonalize procedure. 
Then using eqn.\,(\ref{eqn:eigenu}) and eqn.\,(\ref{eqn:vckmu}), the RG extrapolated value of the 
up-quark mass matrix at the GUT scale is determined

\bea
{\small M^0_u(M_{\rm GUT})}&=&{\small \begin{pmatrix}
0.00973&0.0379-0.00693i&0.0635-0.1671i\\
0.0379+0.00693i&0.2482&2.117+0.000116i\\
0.0635+0.1672i&2.117-0.000116i&51.38\end{pmatrix}}
{\small \mbox{GeV}}.\nonumber \\
\label{MuU}
\eea

\begin{center}
{\bf B. Determination of $M_D$ at GUT scale}
\end{center}

In order to fit the fermion masses at the GUT scale, in addition to the two bidoublets originating 
from two different Higgs representations ${10}_{H_1}$ and ${10}_{H_2}$, we utilize the superheavy 
bidoublet in $\xi (2,2,15) \subset {126}_H$. We will show that even if $\xi$ has to be at the intermediate 
scale $\left( 10^{13}-10^{14} \right)$ GeV to generate the desired value of induced VEV needed for quark-lepton 
mass splitting, the precision gauge coupling unification is unaffected. This fermion mass requires the 
predicted Majorana coupling $f$ to be diagonal and the model predicts experimentally testable RH neutrino 
masses. In the presence of inverse seesaw formula taking into account the small masses and large mixings 
in the LH neutrino sector in the way of fitting the neutrino oscillation data, this diagonal structure 
of $f$ causes no problem. However we show that when we treat the intermediate scale for sub-multiplet to be 
$\xi^{\prime}(2,2,15)$ replacing $\xi (2,2,15)$ but originating from a second Higgs representation 
${126}_H^{\prime}$ which has coupling $f^{\prime}$ to the fermions and all other scalar components 
at the GUT scale, the coupling $f$ and hence $M_N$ can have a general texture, not necessarily diagonal, 
although fermion mass fit needs only $f^{\prime}$ to be diagonal.

The VEV of $\xi(2,2,15)$ is well known for its role in to splitting the quark and lepton masses 
through the Yukawa interaction $f{\bf 16.16.{126}_H^{\dagger}}$ \cite{baburnm93}. It is sometimes 
apprehended, as happens in the presence of only one ${10}_H$, that this new contribution may also upset 
the near equality of $M_u^0 \simeq M_D^0$ at the GUT scale. But in the presence of the two different 
${10}_{H_1}$ and ${10}_{H_2}$ producing the up and down type doublets, the effective theory from the 
$\mu \ge 10$ TeV acts like a nonsupersymmetric two-Higgs doublet model with available large value of $\tan \beta =v_u/v_d$ 
that causes the most desired splitting between the up and down quark mass matrices but ensures $M^0_u \sim M^0_D$. After having achieved 
this splitting a smaller value of of the VEV $v_{\xi}$ is needed to implement fitting of charged fermion
mass matrices without substantially upsetting the near equality of $M_u^0 \simeq M_D^0$ at the GUT scale 
\footnote{It is to be noted that the validity of our estimations of $0\nu\beta\beta$ decay and non-unitarity 
and LFV effects  do not require exact equality 
of $M_u$ and $M_D$ and a relation between them within less than an order of magnitude deviation would suffice 
to make dominant contributions at the TeV scale. But the present  models, either with $G_{224D}$ or $SO(10)$ 
symmetry at the high GUT scale, give the high scale prediction $M_u^0 \sim M_D^0$ up to a good approximation.}.

The formulas for mass matrices at the GUT scale are \cite{Bdev-non,ramlal}
\bea
M_u &=&{G}_u + {F}, ~~M_d ={G}_d + {F},\nonumber\\
M_e &=& {G}_d -3 {F},~~M_D ={G}_u -3 {F}.\label{fmU}
\eea
where $G_k = Y_k \langle {10}^k_H \rangle, k=u,d$ and $F = f v_{\xi}$ leading to

\be
f=\frac{(M_d-M_e)}{4v_{\xi}}. \label{feq}
\ee


Using a charged-lepton diagonal mass basis and eq.\,(\ref{eqn:eigenu}) and eq.\,(\ref{fmU}) we have
\bea
M_e(M_{GUT})&=&{\rm diag}(0.000216,0.0388,0.9620)~{\rm GeV},\nonumber\\
 G_{d,ij} &=& 3F_{ij},~(i \neq j).\label{hdij}
\eea  
\par\noindent{\bf (i) Diagonal structure of RH neutrino mass matrix:}

In refs.\cite{Bdev-non,ramlal} dealing with TeV scale pseudo-Dirac RH neutrinos, a diagonal structure 
of $F$ was assumed with the help of higher dimensional non-renormalizable operators in order to fit 
the charged fermion masses and mixings at the GUT scale. In the present model renormalizable interaction 
of ${126}_H$ is available the diagonal structure of $F$ is a result of utilization of diagonal basis of 
down quarks as well.

This diagonal structure of $f$ would have caused serious problem in fitting the neutrino oscillation data 
if we had a dominant type-II seesaw formula \cite{typeII}, but it causes no problem  in our present model 
where type-II seesaw contribution to light neutrino mass matrix is severely damped out compared 
to inverse seesaw contribution which fits the neutrino oscillation data. Further, the resulting diagonal 
structure of RH neutrino mass matrix that emerges in this model has been widely used in SUSY and non-SUSY 
$SO(10)$ by a large number of authors, and this model creates no anomaly as there are no experimental data 
or constraints which are violated by this diagonal structure.  
 
The quark mixings reflected through the CKM mixing matrix $V_{CKM} = U_L^{\dagger}D_L=U_L$ has been 
parametrized at $\mu=M_Z$ in the down-quark diagonal basis and this mixing matrix has been extrapolated 
to the GUT scale resulting in $V_{CKM}^0\equiv U_L^0$ in eq.\,(\ref{eqn:vckmu}) provided $D_L^0=I$ which 
can hold even at the GUT scale if we use down quark diagonal basis. In that case $M_d (M_{\rm GUT})=M_d^0= {\rm diag}(m_d^0, 
m_s^0, m_b^0)$ which is completely determined by the respective mass eigen values determined by the bottom-up 
approach. Then the second mass relation of eq.\,(\ref{fmU}) gives,
\bea
G_{d_{ij}} &=& -F_{ij},~(i \neq j).\label{hddij} 
\eea 
Now  eq.\,(\ref{hdij}) and eq.\,(\ref{hddij}) are satisfied only if $F_{ij}=0,~(i \neq j)$ i.e, if $F$ 
is diagonal. This is also reflected directly through the eq.\,(\ref{feq}). In other words the diagonality 
of $F$ used in earlier applications of inverse seesaw mechanism in $SO(10)$ \cite{Bdev-non,ramlal} is 
a consequence of utilization of down quark and charged lepton diagonal bases and vice-versa, although 
through non-renormalizable Yukawa interaction. In the present model it shows that even by restricting 
$F$ to its diagonal structure which eliminates at least six additional parameters which would have otherwise 
existed via its non-diagonal elements, the model successfully fits all the charged fermion masses and mixings 
including the Dirac phase of the CKM matrix at the GUT scale. Besides, as shown below, the model predicts 
the RH neutrino masses  accessible to high energy accelerators including LHC. We have relations between 
the diagonal elements which, in turn, determine the diagonal matrices $F$ and $G_d$ completely.
\bea
\mbox{G}_{\rm d,ii} + \mbox{F}_{\rm ii} &=& m^0_i,~(\mbox{i=d,s,b}),\nonumber\\
\mbox{G}_{\rm d,jj} -3 \mbox{F}_{\rm jj}&=& m^0_j,~(j=e,\mu,\tau).\label{hfrel}
\eea
\bea
F&=&{\rm diag}{1\over 4}(m^0_d-m^0_e, m^0_s-m^0_{\mu}, m^0_b-m^0_{\tau}),\nonumber\\
&=&{\rm diag}(2.365\times 10^{-4}, -0.0038, +0.015)~{\rm GeV},\nonumber\\
G_d&=&{\rm diag}{1\over 4}(3m^0_d+m^0_e, 3m^0_s+m^0_{\mu}, 3m^0_b+m^0_{\tau}),\nonumber\\
&=&{\rm diag}(9.2645\times 10^{-4}, 0.027224, 1.00975)~{\rm GeV},\label{FGd}
\eea
where we have used the RG extrapolated values of eq.\,(\ref{eqn:eigenu}). It is clear from the value 
of the mass matrix $F$ in eq.\,(\ref{FGd}) that we need as small a VEV as $v_{\xi}\sim 10$ MeV to 
carry out the fermion mass fits at the GUT scale. In the subsection 6.5.D below we show how the $SO(10)$ 
structure and the Higgs representations given for the symmetry breaking chain of eq.\,(\ref{chain}) clearly 
predicts a VEV $v_{\xi}\sim (10-100)$ MeV consistent with precision gauge coupling unification and 
the fermion mass values discussed in this subsection.

The model ansatz for CKM mixings at the GUT scale matches successfully with those given by $V_{CKM}^0$ 
of eq.\,(\ref{eqn:vckmu}) and, similarly, the model predictions for up quark masses can match with those given 
in eq.\,(\ref{eqn:eigenu}) provided we can identify $M_u$ of eq.\,(\ref{fmU}) with $M_u^0$ of eq.\,(\ref{MuU}). 
This is done by fixing $G_u$=$M_u^0-F$ leading to       
\bea
{\small G_u(M_{GUT})}&=&{\small \begin{pmatrix}
0.00950         & 0.0379-0.00693i  & 0.0635-0.1671i\\
0.0379+0.00693i & 0.2637           & 2.117+0.000116i\\
0.0635+0.1672i  & 2.117-0.000116i  & 51.4436 \end{pmatrix}}\,{\small {\rm GeV}} .\label{GuU}
\eea
Now using eq.\,(\ref{FGd}) and eq.\,(\ref{GuU}) in eq.\,(\ref{fmU}) gives the Dirac
neutrino mass matrix $M_D$ at the GUT scale
\bea
{\small M^0_D(M_{GUT})}&=&{\small \begin{pmatrix}
0.00876          & 0.0380-0.00693i   & 0.0635-0.1672i \\
0.0380+0.00693i  & 0.3102            & 2.118+0.000116i\\
0.0635+0.1672i   & 2.118-0.000116i   & 51.6344
\end{pmatrix}}~{\small {\rm GeV}}.\label{mdU}
\eea 
 
The relation $F=fv_{\xi}={\rm diag}(f_1,f_2,f_3)v_{\xi}$ in eq.(6.15) with $v_{\xi}=10~$ MeV 
gives $(f_1,f_2,f_3)=(0.0236, -0.38, 1.5)$ \footnote{In the context of observable $n-{\bar n}$ oscillation, the value of  $f_1\sim 0.01$ and
quartic coupling $\lambda \sim 1$ need the degenerate  mass of diquark Higgs scalars $M_{\Delta}=5\times 10^4$ GeV. We have checked that
precision gauge coupling unification in the symmetry breaking chain remains unaltered with such mildly tuned value of diquark
Higgs scalars contained in $\Delta_R(1,3,\overline{10})\subset {126}_H$.}. Then the allowed solution to RGEs for gauge coupling
unification with $M_R^0=v_R=5$ TeV determines the RH neutrino masses.
\begin{equation} 
M_{N_1}=115 ~{\rm GeV}, ~M_{N_2}=1.785~{\rm TeV},~M_{N_3}=7.5 ~{\rm }TeV. \label{rhprediction}
\end{equation}

While the first RH neutrino is lighter than the current experimental limit on $Z_R$ boson mass, the second one 
is in-between the $Z_R$ and $W_R$ boson mass limits, but the heaviest one is larger than the $W_R$ mass limit. 
These are expected to provide interesting collider signatures at LHC and future accelerators. This hierarchy 
of the RH neutrino masses has been found to be consistent with lepton-number and lepton flavor violations 
discussed in Sec.\,2, Sec.\,4, and Sec.\,5. 

\begin{figure}[h!]
\begin{center}
\includegraphics[scale=0.39, angle=-90]{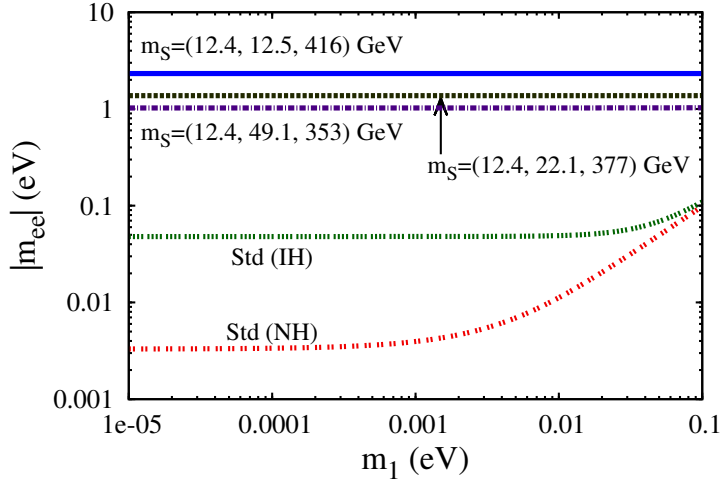}
\caption{Estimations of effective mass parameter for $0\nu\beta\beta$ decay in the $W_L^--W_L^-$ 
         channel with sterile neutrino exchanges shown by top, middle, and bottom horizontal lines. 
         The RH neutrino masses and the Dirac neutrino masses are derived from fermion mass fits 
         and the sterile neutrino masses have been obtained through $M$ values consistent with 
         non-unitarity constraints as described in the text.}
\label{fig:predRHWLWL}
\end{center}
\end{figure}

We estimate effective mass parameters for $0\nu\beta\beta$ decay using this predicted diagonal structure 
of $M_N$ and three sets of constrained $N-S$ mixing matrix $M_i=(40,150, 1810)$ GeV,  $M_i=(40,200, 1720)$ 
GeV, and $M_i=(40,300, 1660)$ GeV corresponding to the three sets of sterile neutrino mass eigen values 
$m_{S_i}=(12.4,12.5, 416)$ GeV,  $m_{S_i}=(12.5, 22.1, 377)$ GeV, and  $m_{S_i}=(12.4, 49, 350)$ GeV, 
respectively. The estimated values of the effective mass parameters in the $W_L-W_L$ channel due to sterile 
neutrino exchanges have been shown in Fig.\ref{fig:predRHWLWL} where the top, middle and the bottom horizontal 
lines represent $m_S^{\rm ee, L}=2.1$ eV, $1.3$ eV , and $1.0$ eV corresponding to the first, second and the 
third set, respectively. Thus the new values are found to be much more dominant compared to the standard 
predictions in this channel. Clearly the Heidelberg-Moscow results can be easily accommodated even for normally 
hierarchical or inverted hierarchical light neutrino masses.

\par\noindent{\bf (ii) General form of RH neutrino mass matrix:-}\\
Although we have shown the emergence of diagonal structure of $M_N$ from the successful fermion mass fits 
at the GUT scale, it is worthwhile to explore as to how this approach may also allow a general structure 
for the Yukawa coupling $f$ of ${126}_H$ and hence the RH neutrino mass matrix while giving a successful 
fit to charged fermion masses at the GUT scale. It is clear from the above discussions that this is not 
possible via renormalizable interaction if the model has only a single ${126}_H$. We introduce a second 
${126}_H^{\prime}$ with its coupling $f^{\prime}$ and all its scalar submultiplets at the GUT-Plank scale 
except for the component $\xi^{\prime}(2,2,15)$ which is tuned to have its mass at the intermediate scale 
$M_{{\xi}^{\prime}} \sim 10^{13}$ GeV-$10^{14}$ GeV.  Also, as before, the VEV of the neutral component of $\Delta_R 
(1,3,{\bar {10}})\subset {126}_H$ is used to contribute to the spontaneous breaking of $G_{2113}\to $ SM, 
but the component $\xi(2,2,15)$ assumes its natural GUT scale mass without the necessity of being at the 
intermediate scale. All our results go through by redefining $F=f^{\prime}v_{{\xi}^{\prime}}$ and 
$v_{{\xi}^{\prime}}=(10-100)$~ MeV is realized in the same way as discussed below in Sec.\,6.5\,D.
In this case the diagonal structure of $f^{\prime}$ gives the same successful fit to charged 
fermion masses and mixings at the GUT scale without affecting the allowed general structure of 
$f$ and $M_N$. Unlike the case (i) with single ${126}_H$ discussed above, as $f_1$ is not constrained to be small,
observable $n-{\bar n}$ oscillation is possible in this case for all diquark Higgs scalar masses
$M_{\Delta}\sim M_C \sim 10^5-10^6$ GeV already
permitted by RGE solutions to precision gauge coupling unification. 

So far we have discussed emergence of dominant $0\nu\beta\beta$ decay rates subject to non-unitarity 
constraints with either a purely diagonal or nearly diagonal $M_N$ matrix with small mixing. To test 
whether such results exist for a general structure, we consider a mass matrix,
\begin{eqnarray}
M_N=\left(
\begin{array}{ccc}
 1853.67+320.545 i & -2165.24-47.9844 i & 2064.69+364.436 i \\
 -2165.24-47.9844 i & 2818.92-210.568 i & -2030.45+245.815 i \\
 2064.69+364.436 i & -2030.45+245.815 i & 4610.57-2.67651 i
\end{array}
\right)
\label{gen_mn} 
\end{eqnarray}
which has the eigen values $M_{N_i}= (115 , 1750, 7500)$ GeV with the same mixings as the LH neutrinos.  
Using eq.\,(\ref{gen_mn}), the non-unitarity constrained $N-S$ mixing matrix $M= {\rm diag}(40, 150, 1810)$ 
GeV, and the derived value of the Dirac neutrino mass matrix from eq.\,(\ref{eq:md_mr0}) leads to the sterile 
neutrino mass eigen values $m_{S_i} \simeq  (1, 48,1500)$ GeV and the resulting effective mass parameters 
in the notations of Sec.\,4- Sec.\,6 are found to be
\bea
m^{\rm ee,L}_{S} &=& 2.5~{\rm eV}, ~m^{\rm ee,R}_{N}= 0.02 ~{\rm eV},~m^{\rm ee,LR}_{\lambda, \eta} = 0.001 {\rm  eV}.\label{predgen}
\eea
Thus, we have shown that the dominant contribution in the $W_L-W_L$ channel due to sterile neutrino exchanges 
estimated using diagonal structure of $M_N$ is also possible with its general structure. Since the form of the 
matrix $M_N$ is not restricted by GUT-scale fermion mass fits, we note that the diagonal forms of $M_N$ used 
in Sec.\,4 and in Fig.\,5-Fig.\,7 and Tables 5-6 to avoid exigency in computation belong to this class which need 
an additional ${126}_H^{\prime}$ within the $SO(10)$ structure.
          
{\bf C.\, Determination of  $M_D(M_{R^0})$ by top-down approach}.

\noindent We use the RGEs in the top-down approach \cite{dp,ramlal,pu} for $M_D$  in the presence 
of $G_{224D}$, $G_{224}$, $G_{2213}$, and $G_{2113}$ to evolve $M_D(M_{GUT})$ to $M_D(M_{R^0})$  through 
$M_D(M_{M_P})$, $M_D(M_{M_C})$ and $M_D(M_{M_R^+})$ and obtain the ansatz given in eqn.\,(\ref{eq:md_with_rge}) 
as
\begin{eqnarray}
\label{eq:md_mr0}
M_D =  \left( \begin{array}{ccc} 
 0.02274          & 0.09891-0.01603i    & 0.1462-0.3859i\\
 0.09891+0.01603i & 0.6319              & 4.884+0.0003034i\\
 0.1462+0.3859i   & 4.884-0.0003034i    & 117.8
\end{array} \right) \text{GeV}\,.
\end{eqnarray}

As can be noted from the determination of running mass eigen values at the high GUT scale 
of the model shown in eqn.\,(\ref{eqn:eigenu}), $b-\tau$ unification is almost perfect, although 
$m^0_{\mu}\simeq 2m^0_{s}$ 
\footnote{While running mass eigenvalues are extrapolated up to the non-SUSY SO(10) unification 
scale in the presence of $G_{2113}$ and $G_{2213}$ intermediate scales \cite{ramlal}, it has been noted that 
at the GUT scale $m^0_b/m^0_{\tau} \simeq 1.3,\, m^0_{\mu}/m^0_s \simeq 2.5$, and $m^0_d/m^0_e \simeq 
4$. Compared to refs. \cite{ramlal,Bdev-non} where a non-renormalizable $dim.\,6$ operator has also been used 
for fermion mass fits at the GUT scale, all the interactions used in this work are renormalizable.}. 
In view of the fact that $G_{224}$ symmetry with unbroken $SU(4)_C$ gauge symmetry is present in this 
model right from $M_C \simeq 10^6$ GeV up to the high GUT scale $M_{GUT}\sim 10^{17.5}$ GeV, the dominance 
of quark lepton symmetry has manifested in the fermion mass relations like $m^0_b\simeq m^0_{\tau} 
\simeq 1.06$ and $M^0_u\simeq M^0_D$ at the GUT scale while making the $SU(4)_C$-breaking effects 
sub-dominant. The bidoublet $\xi(2,2,15) \subset {126}_H$ has been found to make a small contribution 
resulting in the mass matrix $F$ in eq.\,(\ref{FGd}) which plays an important role in our present model. 
The impressive manner in which the underlying quark-lepton symmetry manifests in exhibiting 
$M_u(M_{GUT})\simeq M_D(M_{GUT})$ can be noted from the explicit forms of the two mass matrices 
derived at the GUT scale and shown in eq.\,(\ref{MuU}) and eq.\,(\ref{mdU}).

Thus, the present non-SUSY $SO(10)$ model, having predicted  $M_D$ value given 
in eqn.\,(\ref{eq:md_with_rge}), all our discussions using TeV scale inverse see-saw mechanism 
including neutrinoless double beta decay, non-unitarity effects leading to lepton flavor 
violations, and new CP violating effects discussed in Sec.2-Sec.5, where this mass matrix 
has been used, are also applicable in this GUT model.
\begin{center}
{\bf D.\, Determination of induced vacuum expectation value of $\xi(2,2,15)$ }
\end{center}

\begin{figure}[h]
\centering
\includegraphics[width=11cm, height=13cm, angle=-90]{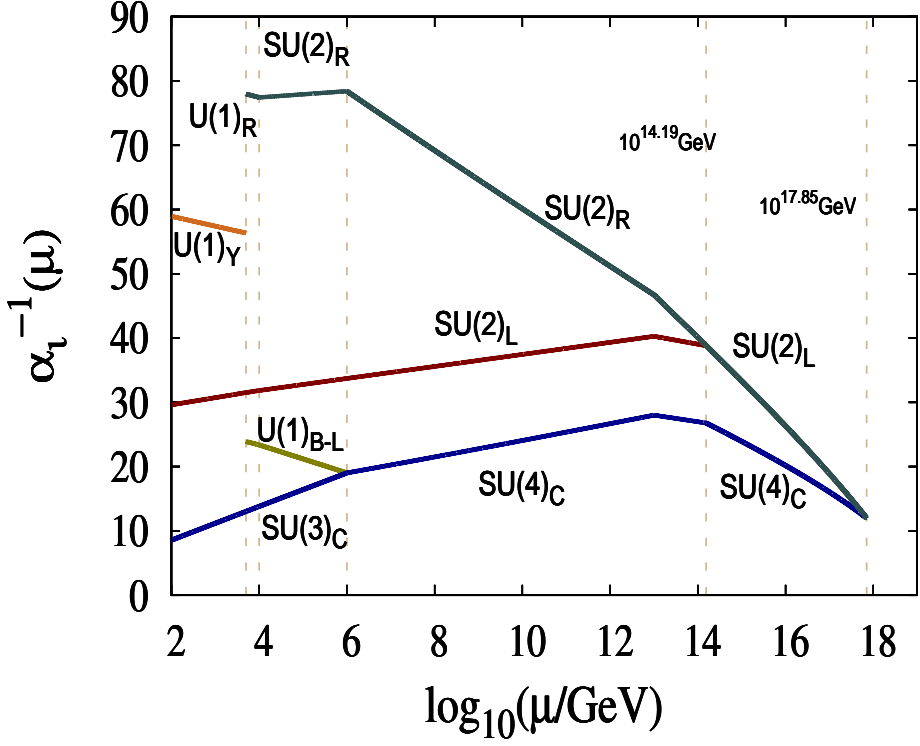}
\caption{Same as Fig.11 but with the scalar sub-multiplet $\xi(2,2,15)$ under Pati-Salam group at $M_\xi=10^{13.2}$ GeV.} 
\label{fig:zeta}
\end{figure}

Now we show how  a small induced VEV $v_{\xi} \sim 10$ MeV of the sub-multiplet $\xi(2,2,15)\subset {126}_H$, 
which has been found to be necessary for fitting the charged fermion masses at the GUT scale, originates 
from the the present $SO(10)$ model. The Higgs representations needed for the symmetry breaking chain permits 
the following term in the Higgs potential
\bea
\lambda_{\xi}\,M^{\prime}\,{210}_H\,{126}_H^{\dagger}\,{10}_H \supset \lambda_{\xi}\,M^{\prime}\,{(2,2,15)}_{126}\,
{(1,1,15)}_{210}\,{(2,2,1)}_{10}\,
\label{vinpot}
\eea
where $M^{\prime}$ is a mass parameter appropriate for trilinear scalar coupling which is naturally of the order 
of the GUT scale $\sim \mathcal{O}(10^{18})$ GeV. For allowed solutions of the mass scales in our model, we have found 
$\langle(1,1,15)\rangle =M_C \simeq 10^6$ GeV, a criteria necessary for observable $n-{\bar n}$ oscillation and rare kaon decay. 
The induced $v_{\xi}$ then turns out to be 
\bea
v_{\xi} = \lambda_{\xi}\,M^{\prime}\,M_C\,v_{ew}/M_{\xi}^2\label{vin}
\eea
\begin{table}
\begin{center}
\begin{tabular}{|c|c|c|c||c|}
\hline
$M_{\xi}$ (GeV) & $M_P$ (GeV) & $M_{\rm GUT}$ (GeV) & $\alpha_G$ & $v_{\xi}$ (MeV) \\ \hline \hline
$10^{13.2}$ & $10^{14.19}$ & $10^{17.83}$ & $0.083$              & 25-250 \\ \hline
$10^{13.4}$ & $10^{14.19}$ & $10^{17.83}$ & $0.076$              & 10-100 \\ \hline
$10^{13.5}$ & $10^{14.19}$ & $10^{17.83}$ & $0.073$              & 7-70 \\ \hline
$10^{13.8}$ & $10^{14.20}$ & $10^{17.82}$ & $0.068$              & 2-16 \\ \hline
$10^{14.0}$ & $10^{14.20}$ & $10^{17.81}$ & $0.065$              & 1-7 \\ \hline
\end{tabular}
\caption{Allowed solutions in the $SO(10)$ symmetry breaking chain shown in eq.\,(6.1) but with the scalar 
component $\xi(2,2,15)\subset {126}_H$ lighter than the GUT scale and consistent with the determination of the 
induced VEV $v_{\xi} \sim (10-100)$ MeV needed to fit charged fermion masses at the GUT scale. For all 
solutions we have fixed $M_R^0=5$ TeV, $M_R^+=10$ TeV, and $M_C=10^6$ GeV.} 
\label{tab:zeta}  
\end{center}
\end{table}

Using $M_C \simeq 10^6$ GeV which is required as model predictions for observable $n-{\bar n}$ oscillation 
and rare kaon decay, and $v_{ew}\sim 100$ GeV, we find that for $\lambda_{\xi}=0.1-1.0$, the eq.\,(\ref{vin}) 
gives the induced VEV $v_{\xi} \simeq (10-100)$ MeV provided $M_{\xi} \sim 10^{13}$ GeV-$10^{14}$ GeV. When $\xi(2,2,15)$ 
is made lighter than the GUT scale having such an intermediate mass, $M_{\xi}=10^{13.4}$ GeV, the precision gauge 
coupling is found to occur as shown in Fig.\ref{fig:zeta} but now with nearly two times larger GUT scale and larger 
GUT fine-structure constant than the minimal case. Our numerical solutions  are shown in Table \ref{tab:zeta} where 
the Parity violating scale is close to the minimal case. It is interesting to note that the precision unification  
with $\xi(2,2,15)\subset {126}_H$ at the intermediate scale is possible without upsetting low mass $W_R$, $Z_R$, 
$M_C$ and other mass scale predictions of the model.  The fermion mass evolutions and the emerging value of $M_D$ 
remain close to the value derived in Sec.\,6.5. The unification pattern and model predictions including GUT-scale 
fermion mass fit are essentially unchanged when the second ${126}_H^{\prime}$ is introduced with its Yukawa coupling 
$f^{\prime}$ and the component $\xi^{\prime}(2,2,15)\subset {126}_H^{\prime}$ at the intermediate scale replacing 
$\xi(2,2,15)\subset {126}_H$ and the latter is assigned its natural GUT scale mass. In this case the mass scales of 
the model give $v_{{\xi}^{\prime}}= (10-100)$ MeV. As the additional threshold contributions to $\sin^2\theta_W$ 
and $M_P$ due to the superheavy components of second ${126}_H^{\prime}$ vanish \cite{pp}, the only change that 
can occur is the GUT-scale threshold effects on $M_{GUT}$. However as the unification scale is close to the Plank 
scale with large proton lifetime prediction, this will not have any additional observable effects.  

Thus, we have shown that the small induced VEV $v_{\xi}$ or $v_{{\xi}^{\prime}}$ needed for GUT scale fit to the 
charged fermion masses and prediction of $M_D$ which is crucial for low-energy estimation of $0\nu\beta\beta$ 
decay rate can be easily derived from the present $SO(10)$ structure. It is possible to have a  diagonal 
structure or a general structure for the RH neutrino mass matrix $M_N$ for which dominant contributions to 
$0\nu\beta\beta$ decay, experimentally accessible LFV decays, and non-unitarity and CP-violating effects have been 
discussed in Sec.\,4 and Sec.\,5.

\subsection{Suppressed induced contribution to $\nu-S$ mixing}
In our model the $\nu-S$ mixing term has been chosen to be vanishingly small in eqn.\,(\ref{eqn:numatrix}). 
However, because of the presence of non-minimal Higgs fields including LH and RH doublets carrying 
$B-L=-1$, triplets carrying $B-L=-2$, two bi-doublets each with $B-L=0$, and Parity odd singlet, it is necessary 
to evaluate if such a term can arise through the induced VEV of $\langle \chi_L\rangle$. We find that without 
taking recourse to any severe fine tuning of parameters, minimization of the scalar potential gives

\begin{equation}
\langle \chi_L \rangle \simeq K \frac{\langle \chi_R \rangle\, v}{M_P}\, ,
\label{indvev}
\end{equation}
 where the ratio of parameters $K={\cal O}(0.1-.01)$ and $M_P \simeq 10^{14}$ GeV. When eqn.\,(\ref{indvev}) 
 is used in the corresponding correction to the light neutrino mass predictions \cite{barr}, $m_{\nu}\simeq 
 M_D\,\frac{\langle \chi_L \rangle}{\langle \chi_R \rangle}$, this gives $m_{{\nu}_{33}} <<  0.001$ eV and 
 negligible contributions to all the three light neutrino masses. With fine-tuning of parameters this contribution 
 can be reduced further. Thus the predictions of the model carried out using eqn.\,(\ref{eqn:numatrix}) are 
 found to hold up to a very good approximation as the small induced contribution $\langle \chi_L \rangle$ 
 does not affect the results substantially. Fine tuning of model parameters would result in   
further reduction of this contribution.

\section{Summary and Conclusion}
In this work we have investigated in detail the prospects of TeV scale left-right gauge theory with high scale 
parity restoration but originating from Pati-Salam or $SO(10)$ grand unified theory  to implement extended 
seesaw mechanism resulting in dominant contributions to $0\nu \beta \beta$ decay  and experimentally accessible 
LFV decays, non-unitarity and CP-violating effects, $n-\bar {n}$ oscillation, and rare kaon decays  while 
preserving the hall-mark of such models to represent all fermion masses and mixings of three generations. 
Consistent with updated values of $\sin^2\theta_W (M_Z), \alpha_S(M_Z)$ and $\alpha (M_Z)$  we have embedded 
the model successfully in non-SUSY Pati-Salam symmetry and  $SO(10)$ GUT predicting low mass $W^{\pm}_R$ and 
$Z_R$ bosons near $1-10$ TeV scale accessible to Large Hadron Collider (LHC) and future accelerators.  We have 
also shown how a novel mechanism  operates to realize grand unification at the GUT-Planck scale through parity 
conserving Pati-Salam symmetry.

The Dirac neutrino mass matrix $M_D$ necessary to estimate lepton number and lepton flavor violating contributions, 
non-unitarity effects as well as  leptonic CP-violation in this model has been explicitly computed using the associated 
renormalization group equations via bottom-up and top-down approaches, and by fitting the quark masses and mixings 
and the charged lepton masses of three generations at the GUT scale. The induced VEV of the Pati-Salam sub-multiplet 
$\xi(2,2,15)$ of ${126}_H\subset SO(10)$ needed to fit the fermion masses at the GUT scale is found to emerge naturally 
within the specified $SO(10)$ structure while safeguarding the precision gauge coupling unification, and values of mass 
scales needed for all experimentally verifiable physical processes as well as the value of the Dirac neutrino mass matrix 
used in our computations. The successful fermion mass fit in the model gives a diagonal structure of RH neutrino 
mass matrix $M_N$ with specific eigen values accessible to accelerator searches. We have also shown how $M_N$ is also 
allowed to be of general form if the model is extended to include an additional  ${126}_H^{\prime}\subset SO(10)$. 

Even though the Dirac neutrino mass matrix is not subdominant but similar to the up quark mass matrix, 
we have shown that low mass $W^\pm_R$ and $Z_R$ bosons, and dominant contributions to $0\nu \beta \beta$ 
decay are in concordance with the neutrino oscillation data for explaining tiny masses of light neutrinos 
which are governed by a gauged inverse seesaw formula near the TeV scale. 

In addition to the Dirac neutrino mass matrix, a major role of the sterile neutrinos,  generically 
required in such inverse seesaw mechanism which has been found in this work is that they give the 
most dominant contributions to the $0\nu \beta \beta$ decay rate with effective mass parameter 
${\bf m^{\rm ee,L}_{S}} \equiv {\bf m_{\mbox{sterile}}^{\rm ee}}\simeq (0.2-2.5)$ eV in the $W^-_L 
-W^-_L$ mediated channel depending upon the allowed range of the sterile neutrino mass eigenvalues. 
In addition, the next dominant contribution in the $W^-_L- W^-_R$ mediated channel due to the exchanges 
of light and heavy RH neutrinos and sterile neutrinos has been also computed. In the $W_R-W_R$ channel 
corresponding to $M_{W_R} \simeq 3-5$ TeV we  find the estimated value of the effective parameter to be about $2$ 
times larger than the standard contribution for NH pattern of light neutrino masses. In addition to 
contributing significantly to the $0\nu \beta \beta$ rate, the quark-lepton symmetric origin of the 
Dirac neutrino mass matrix is also found to play a crucial role in contributing to substantial 
non-unitarity effects leading to enhanced lepton flavor violations and leptonic CP-violation. 
Even with negligible unitarity CP-violation corresponding to the leptonic Dirac phases $\delta_{\rm CP} 
\simeq 0, \pi, 2 \pi$, the models give non-unitarity CP-violating parameter nearly one order larger than the quark sector.

The prediction of $W^{\pm}_R$ and $Z_R$ bosons in this particular non-SUSY $SO(10)$ 
GUT theory is further accompanied by observable $n-\bar{n}$ oscillation with mixing time 
$\tau_{n-\bar{n}} \simeq \left(10^{8}-10^{11}\right)$ secs as well as lepto-quark 
gauge boson mediated rare kaon decay with $\mbox{Br} \left(K_{\rm L} \to \mu \bar{e} 
\right) \simeq \left(10^{-9}-10^{-11} \right)$ accessible to ongoing experiments. 
Another set of important results obtained in these models is noted to include 
non-unitarity effects and branching ratios for LFV decays with $\text{Br}\left(\tau 
\rightarrow e + \gamma \right) = 2.0 \times 10^{-14}, ~ \text{Br}\left(\tau \rightarrow 
\mu + \gamma \right) \leq 2.8 \times 10^{-12} $, and $\text{Br}\left(\mu \rightarrow e + 
\gamma \right) \leq 2.5 \times 10^{-16}$ accessible to ongoing searches. 

We conclude that the experimentally verifiable extended seesaw mechanism in conjunction 
with near TeV scale asymmetric left-right gauge theory accessible to high energy accelerators 
provides a rich structure of weak interaction phenomenology including light neutrino 
masses, neutrinoless double beta decay, non-unitarity effects, and leptonic CP-violation. These can 
originate from  $SO(10)$ grand unified theory or high scale Pati-Salam symmetry with additional verifiable signatures like 
$n-\bar{n}$ oscillation and rare kaon decays. In particular, our finding on sterile neutrino 
mediated dominant $0\nu \beta \beta$ decay rate in the $W^-_L -W^-_L$ channel suggests that 
the Heidelberg-Moscow experimental data could be even consistent with light active neutrinos 
having NH or IH pattern of masses. 
\appendix
\section{APPENDIX}
\label{sec:app-massmix}
The main goal of this section is to derive the masses and mixings for the neutrino sector 
in the extended seesaw mechanism which plays a prime role in determining lepton flavor 
violating processes like $0\nu \beta \beta$ decay rate as well as branching ratios for 
the lepton flavor violating decays. To start with, let us write the complete mass matrix for extended seesaw model in 
flavor basis $\{\nu_L, S_L, N^C_R\}$ as 
\bea
\mathcal{M}_\nu=
\bmt
0     &  0      &  M_D  \\
0     &  \mu_S  &  M^T  \\
M_D^T &  M      &  M_N
\emt
\label{app:numass}
\eea

The flavor basis to mass basis transformation and the diagonalization of the above mass matrix 
is achieved through a unitary matrix
\bea 
& &|\psi\rangle_f=\mathcal{V}\, |\psi\rangle_m \\
&\mbox{or,}&\,\bmt 
\nu_\alpha\\ S_\beta \\ N^C_\gamma
\emt
=
\bmt 
{\cal V}^{\nu\nu}_{\alpha i} & {\cal V}^{\nu{S}}_{\alpha j} & {\cal V}^{\nu {N}}_{\alpha k} \\
{\cal V}^{S\nu}_{\beta i} & {\cal V}^{SS}_{\beta j} & {\cal V}^{SN}_{\beta k} \\
{\cal V}^{N\nu}_{\gamma i} & {\cal V}^{NS}_{\gamma j} & {\cal V}^{NN}_{\gamma k} 
\emt
\bmt 
\hat{\nu}_i \\ \hat{S}_j \\ \hat{N}_k
\emt  
 \label{app:formmix}
\\
&\mbox{and} &\,\mathcal{V}^\dagger \mathcal{M}_\nu \mathcal{V}^*
    =  \hat{\mathcal{M}}_\nu
	 = {\rm diag}\left({ \hat{m}}_{\nu_i};{ \hat{m}}_{{\cal S}_j};{ \hat{m}}_{{\cal N}_k}\right)
	 \label{app:massdiag}
\eea
where subscripts $f, m$ denote for the flavor and mass basis, respectively. Also $\mathcal{M}_\nu$ 
is the mass matrix in flavor basis with $\alpha, \beta, \gamma$ run over three generations of light-neutrinos, 
sterile-neutrinos and right handed heavy-neutrinos in flavor state whereas $\hat{\mathcal{M}}_\nu$ is 
the diagonal mass matrix with $(i,j,k=1,2,3)$ run over corresponding mass states at the sub-eV, GeV 
and TeV scales, respectively. 

Before proceeding to diagonalize the mass matrix, the mass hierarchy $M_N \gg M > M_D, \mu_S$ as well as  
$\mu_S\, M_N < M^2$ has been assumed in our model. The method of complete diagonalization will be carried 
out by two step: ({\bf 1}) the full neutrino mass matrix $\mathcal{M}_\nu$ has to reduced to a block 
diagonalized form as $\mathcal{M}_{\rm \tiny BD}$, ({\bf 2}) this block diagonal form further diagonalized 
to give physical masses of the neutral leptons $\hat{\mathcal{M}}_{\nu}$. 

\subsection{Block diagonalization and determination of $\mathcal{M}_{\rm \tiny BD}$}
We shall follow the parameterization of type given in Ref. \cite{grimus} to determine the form of $\mathcal{W}$. 
In order to evaluate $\mathcal{W}$, let us decompose $\mathcal{W}$ as $\mathcal{W}=\mathcal{W}_1\, 
\mathcal{W}_2$ where $\mathcal{W}_{1}$ and $\mathcal{W}_2$ satisfy 
\bea 
\mathcal{W}_1^\dagger \mathcal{M}_\nu  \mathcal{W}^*_1 = \hat{\mathcal{M}}_{\rm \tiny BD}, \mbox{and}\quad 
\mathcal{W}_2^\dagger \hat{\mathcal{M}}_{\rm \tiny BD}  \mathcal{W}^*_2 = \mathcal{M}_{\rm \tiny BD}
\eea
where $\hat{\mathcal{M}}_{\rm \tiny BD}$, and $\mathcal{M}_{\rm \tiny BD}$ are the intermediate block-diagonal, 
and full block-diagonal mass matrices, respectively,
\bea 
& &\hat{\mathcal{M}}_{\rm \tiny BD} =
\bmt
{\cal M}_{eff}&0\\
0& m_{\cal N}
\emt \\
&\mbox{and}&\, \mathcal{M}_{\rm \tiny BD}
= \bmt m_\nu&0&0\\
0&m_{\cal S}&0\\
0&0&m_{\cal N}
\emt
\eea
\subsubsection{Determination of $\mathcal{W}_1$}
We need to first integrate out the heavy state ($N_R$), being heavier than other mass scales 
in our theory, such that up to the leading order approximation the analytic expressions for 
$\mathcal{W}_1$ is
\bea 
\mathcal{W}_1=\bmt
1-\frac{1}{2}B^*B^T&B^*\\
-B^T&1-\frac{1}{2}B^TB^*
\emt\,, 
\eea 
where the matrix $B$ is $6\times 3$ dimensional and is described as
 
\be 
B^\dagger =M_N^{-1}\left(M^T_D, M^T\right)=(Z^T, Y^T)
\ee
where, $X=M_DM^{-1}$, $Y=M{M_N}^{-1}$, and $Z=M_DM_N^{-1}$ so that 
$Z=X \cdot Y\neq Y\cdot X$ and $y=M^{-1}\mu_S,\, z=M_N^{-1}\mu_S$. 

Therefore, the transformation matrix $\mathcal{W}_1$ can be written purely in terms of dimensionless parameters $Y$ and $Z$
\bea 
\mathcal{W}_1=\bmt
1-\frac{1}{2}ZZ^\dagger & -\frac{1}{2}ZY^\dagger & Z \\
-\frac{1}{2}YZ^\dagger & 1-\frac{1}{2}YY^\dagger & Y \\
-Z^\dagger & -Y^\dagger & 1-\frac{1}{2}(Z^\dagger Z + Y^\dagger Y)
\emt
 \label{app:w1}
\eea
while the light and heavy states can be now written as
\bea
{\cal M}_{eff}&=&\bmt
0&0\\
0&\mu_S
\emt -
\bmt 
M_DM_N^{-1}M^T_D&M_DM_N^{-1}M^T\\
MM_N^{-1}M^T_D& MM_N^{-1}M^T
\emt \\
{ m}_{\cal N}&=&M_N+..
\eea
\subsubsection{Determination of $\mathcal{W}_2$}
From the above discussion, it is quite clear now that the eigenstates $\mathcal{N}_i$ are 
eventually decoupled from others and the remaining mass matrix ${\cal M}_{eff}$ can be 
block diagonalized using another transformation matrix
\bea 
\mathcal{S}^\dagger \mathcal{M}_{\rm eff} \mathcal{S}^* 
      = \bmt 
         m_\nu & 0 \\
         0     & m_{\cal S}
        \emt
\eea
such that
\bea
\mathcal{W}_2 =  \bmt
               \mathcal{S}&0\\
               0&1
                 \emt 
\eea
In a simplified structure 
\bea 
-{\cal M}_{eff}=\bmt 
M_DZ^T & M_DY^T \\
YM_D^T & (MY^T-\mu_S)
\emt
\eea
Under the assumption at the  beginning $Z<<Y$, and of-course $M_D<<M$, this structure is similar to type-(I+II) seesaw. 
Therefore we immediately get the light neutrino masses as
\bea 
m_\nu&=&-M_DZ^T+M_DY^T(MY^T-\mu_S)^{-1}YM^T\nn_D\\
&=&-M_DZ^T+M_DZ^T+M_DM\mu_S(ZY^{-1})^T\nn\\
&=&M_DM^{-1}\mu_S (M_DM^{-1})^T \\
m_{\cal S}&=&\mu_S-MM_N^{-1}M^T
\eea
We see that in addition to $m_{\cal N}$ the $m_{\cal S}$ is also almost diagonal if $M$ and $M_N$ 
are taken to be diagonal. The transformation matrix $S$ is 

\bea 
S=\bmt 
1-\frac{1}{2} A^*A^T&A^*\\
-A^T&1-\frac{1}{2}A^TA^*
\emt
\eea
such that
\bea 
A^\dagger&=&(MY^T-\mu_S)^{-1}YM_D^T\nn\\
&\simeq& (MY^T)^{-1}YM_D^T=X^T\, .
\eea
The $3 \times 3$ block diagonal mixing matrix $\mathcal{W}_2$ has the following form
\bea 
\mathcal{W}_2 
=\bmt 
S & {\bf 0} \\
{\bf 0} & {\bf 1}
\emt = 
\bmt 
1-\frac{1}{2}XX^\dagger &X & 0\\
-X^\dagger & 1-\frac{1}{2}X^\dagger X & 0 \\
0 & 0 & 1
\emt
 \label{app:w2}
\eea
\subsection{Complete diagonalization and physical neutrino masses}
The block diagonal matrices $m_\nu$, $m_{\cal S}$ and $m_{\cal N}$ can further be diagonalized 
to give physical masses for all neutral leptons by a unitary matrix $\mathcal{U}$ as
\bea
\mathcal{U}=\bmt U_\nu & 0 & 0 \cr 0 & U_S & 0 \cr 0 & 0 & U_N \emt.
\label{eq:mixb}
\eea
where the unitary matrices $U_\nu$, $U_{S}$ and $U_{N}$ satisfy
\begin{eqnarray}
U^\dagger_\nu\, m_{\cal\nu}\, U^*_{\nu}  &=& \hat{m}_\nu = 
         \text{diag}\left(m_{\nu_1}, m_{\nu_2}, m_{\nu_3}\right)\, , \nonumber \\ 
U^\dagger_S\, m_{\cal S}\, U^*_{S}  &=& \hat{m}_S = 
         \text{diag}\left(m_{S_1}, m_{S_2}, m_{S_3}\right)\, , \nonumber \\
U^\dagger_N\, m_{\cal N}\, U^*_{N}  &=& \hat{m}_N = 
         \text{diag}\left(m_{N_1}, m_{N_2}, m_{N_3}\right)\,
 \label{app:unit}
\end{eqnarray}
\noindent
With this discussion, the complete mixing matrix is
 \begin{eqnarray}
\mathcal{V}&=&\mathcal{W} \cdot \mathcal{U} =
\mathcal{W}_{1}\cdot \mathcal{W}_{2}\cdot \mathcal{U} \nonumber \\
&=&
\bmt
1-\frac{1}{2}ZZ^\dagger & -\frac{1}{2}ZY^\dagger & Z \\
-\frac{1}{2}YZ^\dagger & 1-\frac{1}{2}YY^\dagger & Y \\
-Z^\dagger & -Y^\dagger & 1-\frac{1}{2}(Z^\dagger Z + Y^\dagger Y)
\emt
\bmt
1-\frac{1}{2}XX^\dagger & X & 0 \\
-X^\dagger  &   1-\frac{1}{2}X^\dagger X & 0 \\
0 & 0 & 1
\emt
\bmt 
U_\nu &0&0\\
0&U_{S}&0\\
0&0&U_{N}
\emt  \nonumber \\
&=&
\bmt
1-\frac{1}{2}XX^\dagger & X-\frac{1}{2}ZY^\dagger & Z \\
-X^\dagger & 1-\frac{1}{2}(X^\dagger X+YY^\dagger) & Y-\frac{1}{2}X^\dagger Z\\
0&-Y^\dagger&1-\frac{1}{2}Y^\dagger Y
\emt
\cdot
\bmt 
U_\nu &0&0\\
0&U_{S}&0\\
0&0&U_{N}
\emt 
 \label{app:mix-extended}
 \end{eqnarray}
 
\newpage 
\section{One- and two-loop beta function coefficients for RG evolution of gauge couplings}
\begin{table*}[htb]
\centering
\begin{tabular}{|c|c|c|c|}
\hline
&&& \\[-4mm]
Group $G_{I}$        & Higgs content    & $\pmb { a_i}$                 & $\pmb{ b_{ij}}$            \\[4mm]
\hline \hline
&&&\\[-4mm]
${\small G_{1_Y2_L3_C}}$                & 
$\begin{array}{l}
\Phi(\frac{1}{2},2,1)_{10}\end{array}$
                               & 
$\bmt 
41/10 \\
-19/6 \\
-7
\emt$ 
                               &  
$\bmt 
199/50   & 27/10    & 44/5  \\
9/10     & 35/6     & 12    \\
11/10    & 9/2      & -26
\emt $ \\[7mm]
\hline 
&&&\\[-4mm]
${\small G_{1_{B-L}1_R 2_L 3_C}}$                                                                  &  
${\small \begin{array}{l}
\Phi_1(0,\frac{1}{2},2,1)_{10}\oplus \Phi_2(0,-\frac{1}{2},2,1)_{10^\prime}\\[2mm]
\Delta_R(-1,1,1,1)_{126} \oplus \chi_R(-\frac{1}{2},\frac{1}{2},1,1)_{16} \end{array}}$   
                                                                                          &
$\bmt 
37/8 \\ 
57/12\\
-3\\
-7
\emt$                                                                                     
                                                                                          &
$\bmt
209/16   & 63/8   & 9/4  & 4 \\
63/8     & 33/4   & 3    & 12\\
3/2      & 1      & 8    & 12\\
1/2      & 3/2    & 9/2  & -26
\emt$ \\[10mm]
\hline 
&&&\\[-4mm]
${\small G_{1_{B-L}2_L 2_R 3_C}}$                 &   
${\small \begin{array}{l}
\Phi_1(0,2,2,1)_{10}\oplus \Phi_2(0,2,2,1)_{10^\prime}\\[2mm]
\Delta_R(-2,1,3,1)_{126} \oplus \chi_R(-1,1,2,1)_{16}\\[2mm]
\Sigma_R(0,1,3,1)_{210}\end{array}}$                    &
$\bmt 
  23/4 \\
  -8/3 \\
  -3/2 \\
  -7
\emt$                                    &
$\bmt
253/8    & 9/2    & 171/4    & 4 \\
3/2      & 37/3   & 6        & 12\\
57/4     & 6      & 263/6    & 12\\
1/2      & 9/2    & 9/2      & -26
\emt$ \\[10mm]
\hline 
&&&\\[-4mm]
${\small G_{2_L2_R4_C}}$                          & 
${\small \begin{array}{l}
\Phi_1(2,2,1)_{10}\oplus \Phi_2(2,2,1)_{10^\prime}\\[2mm]
\Delta_R(1,3,\overline{10})_{126} \oplus \chi_R(1,2,\overline{4})_{16}\\[2mm]
\Sigma_R(1,3,15)_{210} \oplus \sigma^\prime(1,1,15)_{210}
 \end{array}}$
                                         &
$\bmt -8/3 \\
       29/2\\
      -14/3 \emt$                        &
$\bmt 
37/3    & 6       & 45/2   \\
6       & 1103/3  & 1275/2 \\
9/2     & 255/2   & 288
\emt$\\[7mm]
\hline 
&&&\\[-4mm]
${\small G_{2_L 2_R 4_C D}}$                                              &   
${\small \begin{array}{l}
\Phi_1(2,2,1)_{10}\oplus \Phi_2(2,2,1)_{10^\prime}\\[2mm]
\Delta_L(3,1,10)_{126} \oplus \Delta_R(1,3,\overline{10})_{126}\\[2mm]
\chi_L(2,1,4)_{16} \oplus \chi_R(1,2,\overline{4})_{16}\\[2mm]
\Sigma_L(3,1,15)_{210} \oplus \Sigma_R(1,3,15)_{210}\\[2mm]
\sigma^\prime(1,1,15)_{210}  \end{array}}$
                                                                 &
$\bmt 29/3 \\
      29/3 \\
      2/3 \emt$                                                  &
$\bmt 
1103/3  & 6       & 1275/2 \\
6       & 1103/3  & 1275/2 \\
255/2   & 255/2   & 3673/6
\emt$\\[7mm]
\hline
\end{tabular}
\caption{One and two loop beta coefficients for different gauge coupling evolutions described in text
taking the second Higgs doublet at $\mu \geq 10$\,TeV.}
\label{tab:beta_coeff}
\end{table*}
\newpage \noindent
{\bf ACKNOWLEDGEMENT}

\noindent Ram Lal Awasthi acknowledges  hospitality at the Center of Excellence in Theoretical and Mathematical 
Sciences, SOA University where this work was carried out.

\end{document}